%% file: ms.tex
\documentclass[twocolumn,tighten]{aastex62}
\hypersetup{linkcolor=red,citecolor=blue,filecolor=cyan,urlcolor=magenta}

\newcommand{\hdonenine}{\object[HD 19445]{HD~19445}}

\newcommand{\hdsevensix}{\object[HD 76932]{HD~76932}}
\newcommand{\hdeightfour}{\object[HD 84937]{HD~84937}}
\newcommand{\hdninefour}{\object[HD 94028]{HD~94028}}

\newcommand{\hdonefourzero}{\object[HD 140283]{HD~140283}}
\newcommand{\hdonesixzero}{\object[HD 160617]{HD~160617}}

\newcommand{\loggf}{\mbox{$\log gf$}}
\newcommand{\kmsec}{\mbox{km~s$^{\rm -1}$}}
\newcommand{\logg}{\mbox{log~{\it g}}}
\newcommand{\msun}{\mbox{$M_{\odot}$}}
\newcommand{\teff}{\mbox{$T_{\rm eff}$}}
\newcommand{\vt}{\mbox{$v_{\rm t}$}}

\shorttitle{Cu II and Zn II Lines in Late-Type Stars}
\shortauthors{Roederer \& Barklem}
\accepted{for publication in the Astrophysical Journal}

\begin{document}

\title{%
A New Test of Copper and Zinc Abundances in Late-Type Stars 
Using Ultraviolet Cu II and Zn II Lines\footnote{%
Based on observations made with the NASA/ESA 
\textit{Hubble Space Telescope}, 
obtained from the data archive 
at the Space Telescope Science Institute (STScI), which is 
operated by the Association of Universities for 
Research in Astronomy, Inc.\ (AURA) under NASA contract NAS~5-26555.
This work is supported by NASA through grant number AR-15051
and makes use of data from programs 
GO-7348,
GO-8197,
GO-9804,
GO-14161, and
GO-14672.
This research has also made use of the Keck Observatory Archive (KOA), 
which is operated by the W.M.\ Keck Observatory and 
the NASA Exoplanet Science Institute (NExScI), 
under contract with NASA.~
These data are associated with programs
C314Hr,
H6aH, and
H283Hr.
Other data has been obtained from the European Southern Observatory (ESO) 
Science Archive Facility.
These data are associated with programs
65.L-0507(A),
67.D-0439(A), and
080.D-0347(A).
This work has also made use of data collected from
McDonald Observatory of the University of Texas at Austin.%
}
}

\author{Ian U.\ Roederer
}
\affiliation{%
Department of Astronomy, University of Michigan,
1085 S.\ University Ave., Ann Arbor, MI 48109, USA
}
\affiliation{%
Joint Institute for Nuclear Astrophysics -- Center for the
Evolution of the Elements (JINA-CEE), USA
}
\email{Email:\ iur@umich.edu}

\author{Paul S.\ Barklem
}
\affiliation{%
Theoretical Astrophysics,
Department of Physics and Astronomy, Uppsala University,
Box 516, SE-751~20 Uppsala, Sweden
}

\begin{abstract}

We present new abundances derived from
Cu~\textsc{i}, Cu~\textsc{ii}, 
Zn~\textsc{i}, and Zn~\textsc{ii} lines
in six warm (5766~$\leq$~\teff~$\leq$~6427~K),
metal-poor ($-$2.50~$\leq$~[Fe/H]~$\leq -$0.95)
dwarf and subgiant (3.64~$\leq$~\logg~$\leq$~4.44) stars.
These abundances are derived from
archival high-resolution ultraviolet spectra from the
Space Telescope Imaging Spectrograph 
on board the \textit{Hubble Space Telescope} 
and ground-based optical spectra from several observatories.
Ionized Cu and Zn are the majority species, and abundances
derived from Cu~\textsc{ii} and Zn~\textsc{ii} lines
should be largely insensitive to departures from 
local thermodynamic equilibrium (LTE).~
We find good agreement between the [Zn/H] ratios 
derived separately from Zn~\textsc{i} and Zn~\textsc{ii} lines,
suggesting that departures from LTE are, at most, minimal ($\lesssim$~0.1~dex).
We find that the [Cu/H] ratios derived from
Cu~\textsc{ii} lines are 0.36~$\pm$~0.06~dex larger
than those derived from Cu~\textsc{i} lines
in the most metal-poor stars ([Fe/H]~$< -$1.8),
suggesting that LTE underestimates the Cu abundance
derived from Cu~\textsc{i} lines.
The deviations decrease 
in more metal-rich stars.
Our results validate previous 
theoretical non-LTE calculations for both Cu and Zn,
supporting earlier conclusions that
the enhancement of [Zn/Fe] in metal-poor stars
is legitimate, and
the deficiency of [Cu/Fe] in metal-poor stars
may not be as large as previously thought.

\end{abstract}


\section{Introduction}
\label{intro}

One key objective in the study of
late-type stellar abundances is to identify
the limits of calculations made assuming that 
local thermodynamic equilibrium (LTE) 
holds in the line-forming layers
of the atmosphere.
Of particular interest are the elements copper (Cu, $Z =$~29)
and zinc (Zn, $Z =$~30),
two of the heaviest Fe-group elements.
The [Cu/Fe] and [Zn/Fe] ratios
deviate significantly from their Solar ratios
in the most metal-poor stars, which have
presumably been enriched by metals from limited
numbers of earlier nucleosynthesis events.
Specifically, [Cu/Fe] is substantially sub-Solar
in stars with [Fe/H]~$\lesssim -$2,
exhibiting a plateau around
[Cu/Fe]~$\approx -$0.7 to $-$1.0
or a very slight decrease in [Cu/Fe] with decreasing [Fe/H]
(e.g., \citealt{sneden91,mishenina02,bihain04,cayrel04,%
lai08,sobeck08,ishigaki13}).
The [Zn/Fe] ratio increases
to super-Solar ratios
as [Fe/H] decreases
(e.g., in addition to many of the studies listed above,
\citealt{barklem05zn,nissen07,saito09,roederer10,%
hollek11,cohen13,jacobson15,reggiani17}).
It is important that these trends are characterized
reliably so that models of supernova nucleosynthesis and
chemical evolution can be tuned appropriately
(e.g., \citealt{timmes95,goswami00,francois04,kobayashi06,romano07,hirai18}).
The [Zn/H] ratios
in Galactic stars also serve as a metallicity reference 
to compare with damped Lyman-$\alpha$ systems, 
where Zn remains largely in the gas phase
(e.g., \citealt{pettini94,prochaska99,akerman05,rafelski12}).

These stellar abundances are based on LTE radiative transfer calculations
in standard one-dimensional,
hydrostatic model atmospheres.
\citet{takeda05} performed the first non-LTE calculations of
three optical Zn~\textsc{i} lines 
(at 4722, 4810, and 6362~\AA)
to reassess [Zn/Fe] ratios in metal-poor stars
with $-$4~$<$~[Fe/H]~$<$~0.
They found minor non-LTE corrections
to the LTE abundances, usually $<$~0.1~dex,
in the sense that LTE underestimates the abundance.
The corrections were largest in the most metal-poor stars,
mildly exaggerating the observed [Zn/Fe] enhancement
in low-metallicity stars.

\citet{yan15} used the Cu model atom developed by
\citet{shi14} to perform non-LTE calculations for three 
high-excitation optical Cu~\textsc{i} lines 
(at 5105, 5218, and 5782~\AA)
in late-type stars with $-$1.6~$<$~[Fe/H]~$< -$0.1.
Each of these lines behaved slightly differently
when calculated in non-LTE,
but all showed significant positive corrections
to the LTE abundances.
The corrections were smallest for the metal-rich stars, 
$\approx +$0.03~dex,
and increased to nearly $\approx +$0.20~dex for the 
most metal-poor star.
\citet{yan16} obtained similar results when they 
applied non-LTE calculations to rederive [Cu/Fe] ratios
in the sample of \citet{nissen10}.
\citet{andrievsky18} performed non-LTE calculations for
Cu~\textsc{i} lines in dwarf and giant stars spanning
$-$4.2~$<$~[Fe/H]~$< -$1.4.
For the lowest-metallicity stars in their sample,
which were cool giants, only the Cu~\textsc{i} resonance lines
at 3247 and 3273~\AA\ were available.
The non-LTE corrections were substantially larger in these cases,
ranging from $\approx +$0.5 to $+$1.2~dex in 
seven stars with [Fe/H]~$< -$2.
That study was also able to resolve the issue of
discrepant abundances derived from the Cu~\textsc{i}
lines at 3273 and 5105~\AA,
an effect noticed in a metal-poor giant star
by \citet{bonifacio10}.
\citeauthor{yan15}\ and \citeauthor{andrievsky18}\
had one line (Cu~\textsc{i} $\lambda$5105~\AA) in 
one star (\hdninefour) in common, and
they derived different non-LTE corrections to the Cu abundance,
$+$0.17~dex and $+$0.48~dex, respectively.
\citeauthor{andrievsky18}\ attributed this discrepancy to
differences in the adopted model atoms.

One way to test the reliability of the non-LTE calculations for 
Cu and Zn is to derive abundances from Cu~\textsc{ii} and
Zn~\textsc{ii} lines, which arise from the majority species
of each atom and should be relatively immune to non-LTE effects.
The challenge is that Cu~\textsc{ii} and Zn~\textsc{ii} lines
are only found in the ultraviolet (UV) part of the spectrum
of late-type stars, far below the atmospheric cutoff.
Acquiring such spectra requires access to an echelle spectrograph
in space.
\citet{roederer12} made the first such attempts,
using a high-resolution 
($R \equiv \lambda/\Delta\lambda \sim$~114,000) UV spectrum 
of the metal-poor subgiant star \hdonesixzero.
One UV Zn~\textsc{ii} line gave concordant results with
several optical Zn~\textsc{i} lines, but 
the UV Cu~\textsc{ii} lines predicted abundances about a factor of two
higher than the optical Cu~\textsc{i} lines.
\citet{roederer14d} conducted a study of
high-resolution (but less so, $R \sim$~30,000) UV spectra
of two metal-poor giants, finding similar results,
but with large ($\approx$~0.3--0.5~dex) uncertainties
that could mask substantial differences.
\citet{roederer16ipro} also studied Cu~\textsc{i} and \textsc{ii} 
and Zn~\textsc{i} and \textsc{ii} lines
in the metal-poor main sequence star \hdninefour,
reporting agreement between the neutral and ionized species,
but with combined uncertainties $\approx$~0.20--0.25~dex.
\citet{andrievsky18} analyzed high-resolution UV spectra
of two warm, metal-poor stars, finding that their
Cu~\textsc{ii} lines yielded LTE abundances consistent with
non-LTE calculations for optical Cu~\textsc{i} lines.
For completeness, we note that 
Zn can only be detected via
resonance lines of Zn~\textsc{i} and Zn~\textsc{ii} in the UV
in the most metal-poor stars
\citep{roederer16}.

Here, we present a fresh reanalysis of Cu~\textsc{i} and \textsc{ii} 
and Zn~\textsc{i} and Zn~\textsc{ii} lines in
six stars with high-quality UV spectra.
We minimize systematic effects by
(1) selecting stars with a limited range of effective temperatures (\teff)
and surface gravities (\logg);
(2) rederiving stellar parameters and metallicities by 
a consistent method; and
(3) analyzing the lines in a consistent manner
with the same atomic line data,
grid of model atmospheres, and line analysis code.
This enables us to assess the reliability of
LTE calculations for Cu~\textsc{i} and Zn~\textsc{i} lines
in warm, metal-poor stars
and check the reliability of the non-LTE calculations.

Throughout this work,
we adopt the standard nomenclature 
for elemental abundances and ratios.
For element X, the absolute abundance is defined
as the number of X atoms per 10$^{12}$ H atoms,
$\log\epsilon$(X)~$\equiv \log_{10}(N_{\rm X}/N_{\rm H})+$12.0.
For elements X and Y, the abundance ratio relative to the
Solar ratio is defined as
[X/Y]~$\equiv \log_{10} (N_{\rm X}/N_{\rm Y}) -
\log_{10} (N_{\rm X}/N_{\rm Y})_{\odot}$.
We adopt the Solar photospheric abundances of \citet{asplund09}:\
$\log\epsilon$(Fe)~$=$~7.50,
$\log\epsilon$(Cu)~$=$~4.19, and
$\log\epsilon$(Zn)~$=$~4.56.
By convention,
abundances or ratios denoted with the ionization state
are understood to be
the total elemental abundance as derived from transitions of
that particular ionization state 
after Saha ionization corrections have been applied.

\section{Observations}
\label{obs}

We download UV spectra from the
Mikulski Archive for Space Telescopes (MAST).~
These spectra were collected using the
Space Telescope Imaging Spectrograph
(STIS; \citealt{kimble98,woodgate98}) on board the
\textit{Hubble Space Telescope} (\textit{HST}).~
We consider a star for 
inclusion in the sample if extant 
$R \sim$~30,000 or $R \sim$~114,000
spectra cover at least
2037~$\leq \lambda \leq$~2127~\AA,
where the Cu~\textsc{ii} and Zn~\textsc{ii} lines are found.
Inspection of the data reveals that
stars that are warm
(\teff~$>$~5600~K) and
metal-poor 
([Fe/H]~$< -$1.0 or so)
with $R \sim$~114,000 spectra
provide tolerable levels of line blending 
and a reliable estimate of the local continuum.
The STIS $R \sim$~114,000 
observations were made using the E230H echelle grating,
the 0\farcs09~$\times$~0\farcs2 slit, and
the NUV Multianode Microchannel Array 
detector.
The six stars meeting these criteria are listed in Table~\ref{obstab}
along with the instrument, program identification (ID) number, datasets, 
principle investigator (PI), resolving power, and
signal-to-noise (S/N) ratio per pixel 
in the continuum at a reference wavelength.

\input{tab1}

We supplement the UV spectra with
high-resolution optical spectra
downloaded from online archives,
including the 
European Southern Observatory (ESO) Science Archive Facility
and the Keck Observatory Archives (KOA).~
These data were collected with the
Ultraviolet and Visual Echelle Spectrograph (UVES; \citealt{dekker00})
on the Very Large Telescope,
the High Accuracy Radial velocity Planet Searcher (HARPS; \citealt{mayor03}) 
on the ESO 3.6~m Telescope at La Silla, and
the High Resolution Echelle Spectrometer (HIRES; \citealt{vogt94})
on the Keck~I Telescope.
Table~\ref{obstab} lists the instrument,
program ID number, PI, resolving power, and S/N ratio
associated with these data.
An optical spectrum of one star
was collected previously using the
Robert G.\ Tull Coud\'{e} Spectrograph \citep{tull95} at the 
Harlan J.\ Smith Telescope at McDonald Observatory
(see details in \citealt{roederer14}).
The McDonald Observatory does not have an online archive,
and this spectrum may be obtained by contacting IUR directly.

\section{Stellar Parameters}
\label{params}

\subsection{Fe Lines}
\label{linelist}

We compile a list of Fe~\textsc{i} lines 
with reliable oscillator strengths from
the National Institute of Standards and Technology (NIST)
Atomic Spectral Database (ASD; \citealt{kramida17})
with grades C or better ($\leq$~25\% uncertainty, or 0.12~dex).
These \loggf\ values mainly come from work by
\citet{obrian91}.
We supplement this list with results from recent laboratory studies
by \citet{denhartog14} and \citet{ruffoni14},
whose uncertainties are generally reliable to better than 5\% (0.02~dex).
We discard any Fe~\textsc{i} lines with lower excitation potential
(E.P.)\ levels less than 1.2~eV, because previous studies have shown that
these lines may yield higher-than-average abundances 
in metal-poor dwarfs and giants
(e.g., \citealt{cayrel04,cohen08,cohen13,lai08,bergemann12}).

We adopt \loggf\ values for Fe~\textsc{ii} lines from \citet{melendez09}.
Their values for the Fe~\textsc{ii} lines in our study
were obtained by 
renormalizing relative theoretical \loggf\ values within a given multiplet
to an absolute scale set by laboratory measurements of the
upper level lifetime.
We also discuss consideration of
an alternative set of Fe~\textsc{ii} \loggf\ values from NIST,
with \loggf\ grades C or better,
in Section~\ref{modelatmosphere}.

We measure equivalent widths (EWs) of Fe lines
using a semi-automatic 
routine that fits Voigt or Gaussian line profiles to 
continuum-normalized spectra
\citep{roederer14}.
We visually inspect each line, and
any line determined to be
blended, suffer from uncertain
continuum placement, or otherwise compromised
is discarded from consideration.
We retain 97--152 Fe~\textsc{i} lines and
12--17 Fe~\textsc{ii} lines,
spanning 3724--7749~\AA,
and these lines and our EW measurements
are reported in Table~\ref{fetab}.

\input{tab2-stub}

\subsection{Model Atmosphere Parameters}
\label{modelatmosphere}

All stars in the sample are relatively bright and nearby,
so broadband photometry is readily available.
Optical and near-infrared magnitudes 
for these stars are available in catalogs by
\citet{ducati02} and \citet{munari14} 
for Johnson $B$ and $V$,
the Two Micron All Sky Survey (2MASS; \citealt{cutri03}) 
for $J$, $H$, and $K$;
and
\citet{paunzen15} for Str\"{o}mgren-Crawford $b$ and $y$.
We construct six colors
with calibrated color-\teff\ relations
($B-V$, $V-J$, $V-H$, $V-K$, $J-K$, and $b-y$).
We adopt reddening estimates from \citet{casagrande11}.
These stars are located in the foreground 
of the reddening layer,
so $E(B-V) <$~0.005 for all six stars
(see also \citealt{melendez10}).
We deredden colors according to the
extinction coefficients of \citet{mccall04}.

Parallaxes calculated from the  
\textit{Tycho-2 Gaia} Astrometric Solution
(\textit{Gaia} Data Release 1; \citealt{gaia16}) 
are available for 5~stars in the sample.
A parallax measurement for the other star (\hdeightfour) is available
from the \textit{Hipparcos} re-reduction by \citet{vanleeuwen07}.
For the other five stars,
the \textit{Hipparcos} and \textit{Tycho-2/Gaia} parallaxes
agree to better than their 1.6$\sigma$ combined uncertainties.

We adopt \teff\ values calculated from the
\citet{casagrande10}
metallicity-dependent color-\teff\ calibrations.
For each star, the final \teff\ is determined 
using a Monte Carlo resampling method.
We draw $10^{4}$ 
samples from each input parameter
(magnitudes, reddening, and metallicity).
We assume Gaussian uncertainties, including
conservative minimum uncertainties of 0.02~mag in magnitude,
0.02~mag in $E(B-V)$, and 0.3~dex in metallicity.
Each calculation is self-consistent and uses
the same set of input draws,
and we adopt
the median of the final distribution as \teff.
We estimate the uncertainty
as the quadrature sum of the
standard deviation of the \citeauthor{casagrande10}\
\teff\ distribution and the
standard deviation of the \teff\ values
predicted by the \citeauthor{casagrande10},
\citet{alonso99}, and \citet{ramirez05}
color-\teff\ calibrations.
We report these values in Table~\ref{paramtab}.

\input{tab3}

We calculate \logg\
from fundamental relations:\
\begin{eqnarray}
\log g = 4 \log \teff + \log(M/\msun) - 10.61 + 0.4(BC_{V}
  \nonumber\\
  + V - 5\log d + 5 - 3.1 E(B-V) - M_{\rm bol,\odot}).
\end{eqnarray}
Here, $M$ is the mass of the star \citep{casagrande10},
$BC_{V}$ is the bolometric correction in the $V$ band
\citep{casagrande14},
$V$ is the apparent magnitude, and
$d$ is the distance in pc, which is
calculated from the parallax measurement.
$M_{\rm bol,\odot}$ is the Solar bolometric magnitude, 4.75, and
the constant 10.61 is calculated from the Solar constants
$\log \teff_{\odot} =$~3.7617 and $\log g_{\odot} =$~4.438.
We draw
$10^{4}$ samples from each
of these input parameters.
The median of these calculations gives the \logg\ value,
and their standard deviation gives the uncertainty.
We report these values in Table~\ref{paramtab}.
The \teff\ and \logg\ values for these stars place them 
on the main sequence or slightly evolved beyond the 
turnoff point.

We interpolate model atmospheres from the $\alpha$-enhanced
ATLAS9 grid of models \citep{castelli03},
using an interpolation code provided by
A.\ McWilliam (2009, private communication).  
We derive Fe abundances 
using a recent version of the LTE
line analysis software MOOG
(\citealt{sneden73}; 2017 version).
Rayleigh scattering, which affects 
the continuous opacity at shorter wavelengths,
is treated as isotropic, coherent scattering
as described in \citet{sobeck11}.
We adopt 
damping constants for collisional broadening
with neutral hydrogen from \citet{Barklem2000d} 
and \citet{Barklem2005c}, when available,
otherwise
we adopt the standard \citet{unsold55} recipe,
except as noted in Section~\ref{cuznlinesub}.

We iteratively determine 
the microturbulent velocity, \vt, and model metallicity, [M/H].~
Lines yielding an abundance more than 
0.4~dex from the mean are culled.
Convergence is reached when there is no dependence
between line strength and abundance derived from
Fe~\textsc{i} lines, and when the input
[M/H] agrees with the derived [Fe/H] ratios
to within 0.1~dex.
Metallicity uncertainties are estimated by drawing 250 samples of
model parameters and EWs from normal distributions,
assuming a 5\% uncertainty in each EW measurement.
We recompute the abundance for each set, and the
standard deviation of the results is adopted as the uncertainty.
These values are also listed in Table~\ref{paramtab}.

The agreement between
Fe abundances derived
from Fe~\textsc{i} and Fe~\textsc{ii} lines is superb.
The mean difference, in the sense of 
[Fe~\textsc{ii}/H]$-$[Fe~\textsc{i}/H], is  
$+$0.02~$\pm$~0.02 ($\sigma =$~0.05).
Our method of deriving model parameters and Fe abundances
does not enforce Boltzmann excitation equilibrium or
Saha ionization equilibrium,
so this agreement does not result by construction.
Instead, when Fe~\textsc{i} lines with E.P.\ $<$~1.2~eV are excluded,
departures from LTE
in neutral Fe are relatively small for the stars in this sample.
This matches predictions from non-LTE calculations, which show that
most Fe~\textsc{ii} lines are essentially in LTE and
the amount of Fe~\textsc{i} over-ionization has a minor impact
($\lesssim$~0.1~dex) in warm, metal-poor stars
(e.g., \citealt{mashonkina11,bergemann12,sitnova15}).

However, 
the choice of which \loggf\ scale to adopt for Fe~\textsc{ii} lines
plays a significant role in determining the level of agreement.
The NIST \loggf\ values for Fe~\textsc{ii}
decrease the [Fe~\textsc{ii}/H] ratios by
0.09--0.15~dex,
and the average [Fe~\textsc{ii}/H]$-$[Fe~\textsc{i}/H] 
would be $-$0.10~$\pm$~0.03 ($\sigma =$~0.07).
The NIST \loggf\ values for the Fe~\textsc{ii} lines in our study
have typical uncertainties of 0.12~dex,
so the differences between the NIST and \citet{melendez09}
\loggf\ scales are within the stated uncertainties.
We trust that future laboratory work may help to
improve these values.
In the meantime,
we note that the key results of the present study, the
[Cu~\textsc{ii}/Cu~\textsc{i}] and
[Zn~\textsc{ii}/Zn~\textsc{i}] ratios,
are calculated independently from the [Fe/H] ratios.

\subsection{Comparison with Previous Results}
\label{previousparams}

We now compare our \teff,
\logg, and [Fe/H] values with those from previous work.
We select studies for comparison that have at least
three stars in common with our sample.
We also require that the previous study
derived at least one of the ratios [Cu/Fe] or [Zn/Fe]
for each star,
which we use for further comparisons in
Section~\ref{previousabund}.
Six studies meet these criteria:\
\citet{sneden91},
\citet{mishenina01} and \citet{mishenina02} (considered as one),
\citet{bihain04},
\citet{nissen07},
\citet{roederer12c}, and
\citet{bensby14}.

For \hdonenine, \hdsevensix, \hdeightfour,
\hdninefour, \hdonefourzero, and \hdonesixzero,
this results in a total number of stellar parameter comparisons
of 3, 5, 5, 4, 6, and 2~stars, respectively.
We find no evidence that our values differ substantially
from previous work.
The differences between our \teff, \logg, and [Fe/H] values
all agree to better than 2 standard deviations of the mean 
of those from previous studies for all stars except \hdonesixzero.
Only two previous studies meet our criteria for \hdonesixzero,
\citet{nissen07} and \citet{bensby14}.
Our \teff\ value is in good agreement with previous work.
The \logg\ and [Fe/H] values of the two previous studies agree to within
a few hundredths of a dex, producing small standard deviations.
Our \logg\ value is higher than their mean by only 0.06~dex, 
and our [Fe/H] value is lower than their mean by only 0.12~dex.
Both of these are well within the combined uncertainties,
and we regard the stellar parameters of \hdonesixzero\ as
also being in agreement with previous work.

\section{Copper and Zinc Abundance Analysis}
\label{cuznlines}

\subsection{Cu and Zn Lines}
\label{cuznlinesub}

Several Cu~\textsc{i} and Zn~\textsc{i} lines are 
found in the optical part of the spectrum,
while the Cu~\textsc{ii} and Zn~\textsc{ii} lines
are only found in the UV.~
Table~\ref{atomictab} lists the wavelengths, 
E.P., and \loggf\ values for the
lines used in this study.

\input{tab4}

We adopt the \loggf\ values for the Cu~\textsc{i}
lines at 3247.54 and 3273.96~\AA\ from NIST.~
Both values have grades of AA, indicating uncertainties
$\leq$~1\% (0.004~dex).
The \loggf\ values of the Cu~\textsc{i} lines at 5105.54 and
5218.20~\AA\
each have C+ grades ($<$~18\%, 0.09~dex), 
according to NIST.~
Additional Cu~\textsc{i} lines at 
5700.24 and 5782.13~\AA\
are not reliably detected
in our spectra.

\citet{roederer12} evaluated the quality of oscillator strengths available
for seven Cu~\textsc{ii} lines.
\citet{kramida17cu} have released an updated critical compilation of
Cu~\textsc{ii} oscillator strengths, 
which are included in the most recent NIST ASD update.
The two sets of oscillator strengths agree to within their
stated uncertainties, and we simply adopt the values
recommended by \citeauthor{roederer12}.
Four of these lines, listed in Table~\ref{atomictab}, 
are detected in most of our spectra.

\citet{roederer12} also evaluated the oscillator strengths for
Zn~\textsc{i}
lines commonly used in abundance analyses of late-type stars, 
and we adopt those recommended values.
Their uncertainties range from 0.02 to 0.08~dex.
We adopt the \loggf\ value for the Zn~\textsc{i} line at 2138.56~\AA\
from NIST.

We adopt the \loggf\ value for the Zn~\textsc{ii} resonance
line at 2062.00~\AA\ from the work of \citet{bergeson93},
who estimated an uncertainty of 0.03~dex.
The other Zn~\textsc{ii} resonance line at 2025.48~\AA\ is
too blended to be useful as an abundance indicator.

The Zn~\textsc{ii} resonance line 
($4s \, ^2\mathrm{S} - 4p \, ^2\mathrm{P}^\mathrm{o}$)
at 2062.00~\AA\ 
is saturated in all stars in the sample,
and its abundance sensitivity is expressed through its damping wings.
We calculate the broadening of this line
with neutral hydrogen atoms using the 
Anstee, Barklem and O'Mara (ABO) theory, 
originally developed for neutral species
\citep[e.g.,][]{anstee_investigation_1991,Anstee1995,Barklem1997a}, 
and extended to singly-ionized species by \citet{Barklem1998a}.
The neutral and singly-ionized cases
differ in that an approximation originally due to
\citet{unsold55} is not expected to be generally valid for ions.
Specifically, an energy debt parameter, $E_p$, appearing in the ABO theory
is approximated to $E_p=-4/9$ atomic units for neutrals, 
but for ions it is usually calculated in detail
on a line-by-line basis.
However, such calculations require good knowledge 
of the fundamental atomic data 
(wavelengths, oscillator strengths) for transitions in the ion.  
Lack of such data for Zn~\textsc{ii} means that 
a detailed calculation of $E_p$ is not possible.
Thus, we calculate the broadening 
with the \citeauthor{unsold55}\ approximation value, 
$E_p=-4/9$ atomic units.
Within this approximation we obtain the broadening cross section 
at relative velocity $v=10^4$~m/s to be 161 atomic units, 
with a velocity dependence parameter $\alpha = 0.30$ 
\citep[see][]{Anstee1995}.
The log of the line width (full width at half maximum) per perturber at 
10$^{4}$~K is $-$7.9394 rad~s$^{-1}$~cm$^{3}$.
Test calculations reveal that the resonance line is not 
terribly sensitive to this assumption.  
Doubling or halving the the value of $E_p$, 
which roughly spans the range of values seen in other atoms 
\citep[][]{Barklem1998a,Barklem2000d,Barklem2005c}, 
leads to variation of the cross section by at most 25\%, 
and this is perhaps indicative of the error in the calculated value.  
We note that for Fe~\textsc{ii}, $E_p=-4/9$ atomic units 
is in fact a good approximation for the vast majority of levels 
\citep[see Figure~2 of][]{Barklem2005c}.

Cu is an odd-$Z$ element with two $I =$~3/2 isotopes,
$^{63}$Cu and $^{65}$Cu,
giving rise to hyperfine structure (HFS) and isotope shifts (IS)
that may impact the observed line profiles and derived abundances.
We adopt a Solar isotope mixture for Cu in our syntheses.
For Cu~\textsc{i}, we adopt the 
HFS/IS patterns given by the \citet{kurucz11} database.
No modern measurements of the IS,
magnetic dipole, or electric quadrupole
are available for the levels connected by 
our Cu~\textsc{ii} transitions of interest.
\citet{roederer12} assembled trial
HFS/IS patterns 
from early work by \citet{elbel63} and \citet{elbel69}
for the Cu~\textsc{ii} line at 2126~\AA\
to gauge the potential impact of neglecting the HFS/IS
on the derived abundances, and
we repeat this exercise here.
For most stars, there is no discernible impact
at the 0.01~dex level.
Only in the most metal-rich stars, 
\hdsevensix\ and \hdninefour,
do the HFS/IS component patterns
result in a mild abundance decrease,
by 0.04 and 0.02~dex, respectively.
Data are not available for the other Cu~\textsc{ii} lines of interest here.
We conclude that the derived Cu~\textsc{ii} abundances
in \hdsevensix\ and \hdninefour\ could be overestimated 
by a few hundredths of a dex.
No HFS/IS component patterns are considered for any
Zn~\textsc{i} or \textsc{ii} line.

\subsection{Abundances}

We derive abundances for
all Cu and Zn lines 
by comparing the observed line profiles
to MOOG synthetic spectra with varied Cu or Zn abundances.
We generate line lists using a modified version of
the \textit{linemake} code
(C.\ Sneden 2017, private communication),
which includes additional updates for UV transitions
relevant to late-type stellar spectra.
This code replaces the \citet{kurucz11} line list entries with
wavelengths, \loggf\ values, molecular lines,
and HFS/IS component patterns from modern experimental work.
\citet{lawler09,lawler17} summarize the methods and
limitations of these laboratory efforts.

Abundances derived from individual Cu~\textsc{i} and
\textsc{ii} lines are reported in Table~\ref{cutab},
and 
abundances derived from individual Zn~\textsc{i} and
\textsc{ii} lines are reported in Table~\ref{zntab}.
Table~\ref{cuabundtab} lists 
the mean Cu abundance ratios,
and 
Table~\ref{znabundtab} lists 
the mean Zn abundance ratios.
The statistical uncertainty estimates
account for the impact of noise, 
uncertainties in the continuum placement, 
minor blends, 
and \loggf\ values.
The systematic uncertainty estimates
account for the impact of 
uncertainties in the model atmosphere parameters.
We estimate the systematic uncertainties
by recomputing the
Cu or Zn abundances in a series of 250 trials,
each with a different set of 
approximate Cu or Zn EWs and
model atmosphere parameters.

\input{tab5}

\input{tab6}

\input{tab7}

\input{tab8}

\begin{figure*}
\includegraphics[angle=0,width=3.4in]{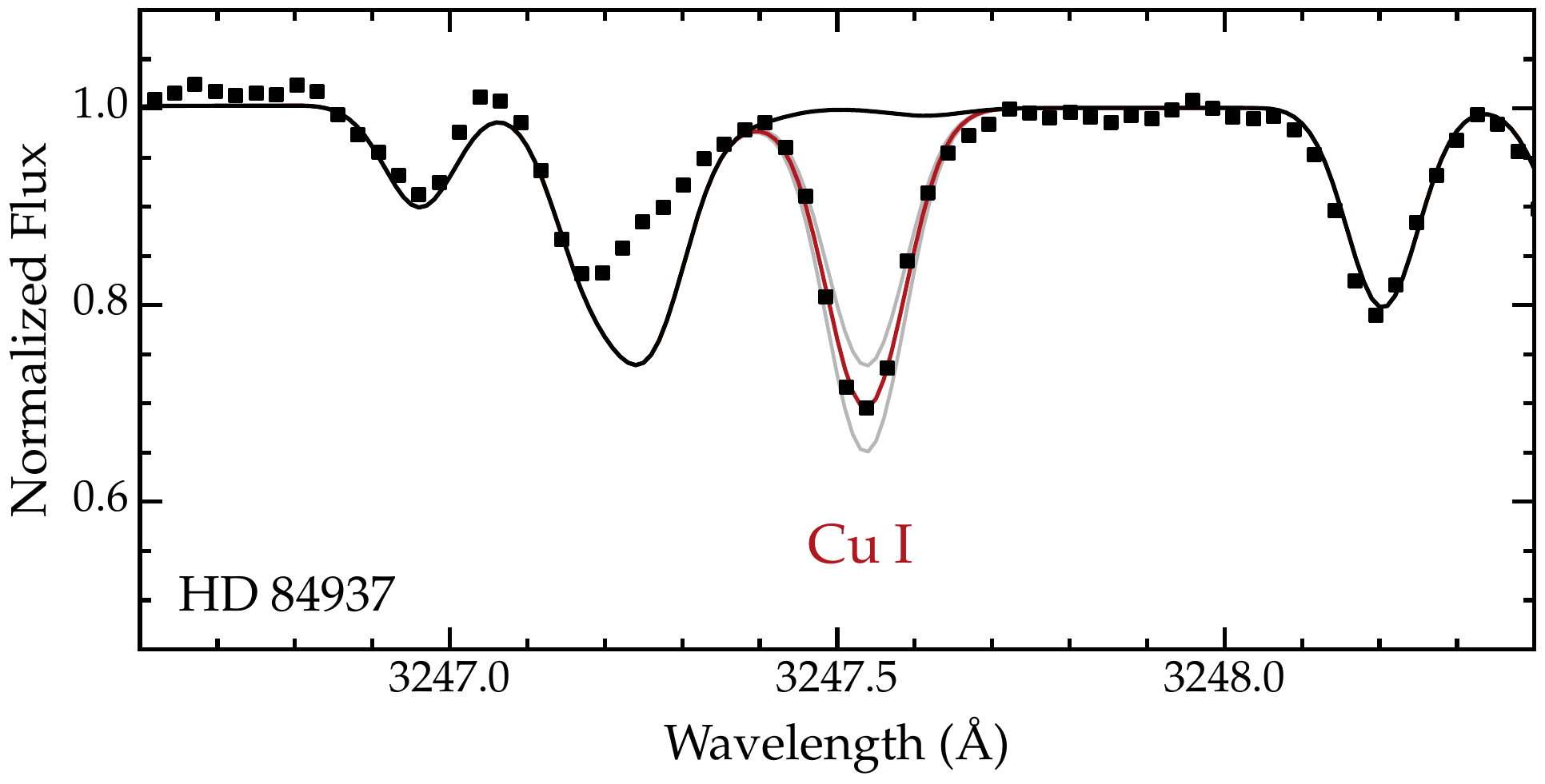} 
\hspace*{0.1in}
\includegraphics[angle=0,width=3.4in]{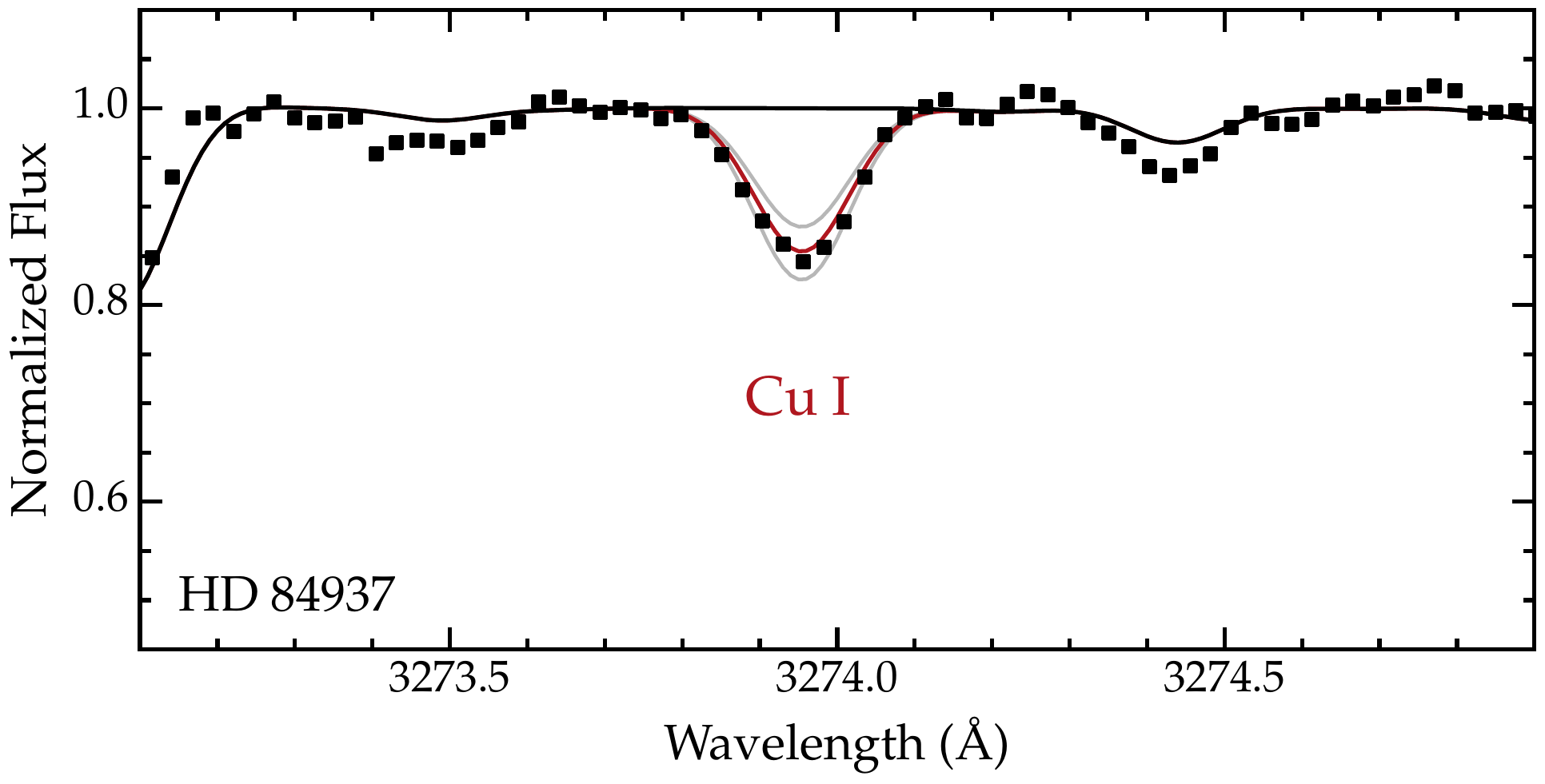} \\
\includegraphics[angle=0,width=3.4in]{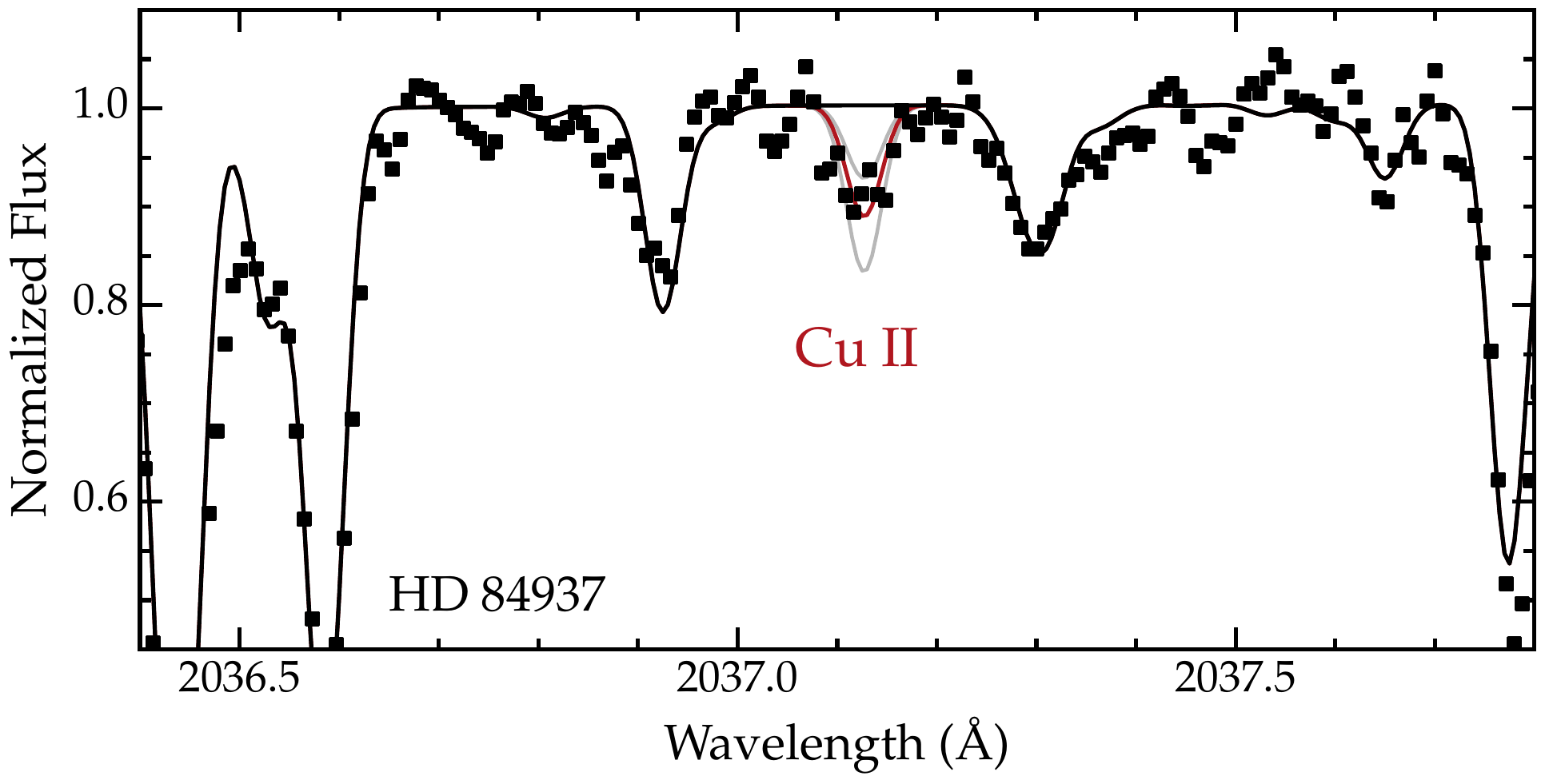} 
\hspace*{0.1in}
\includegraphics[angle=0,width=3.4in]{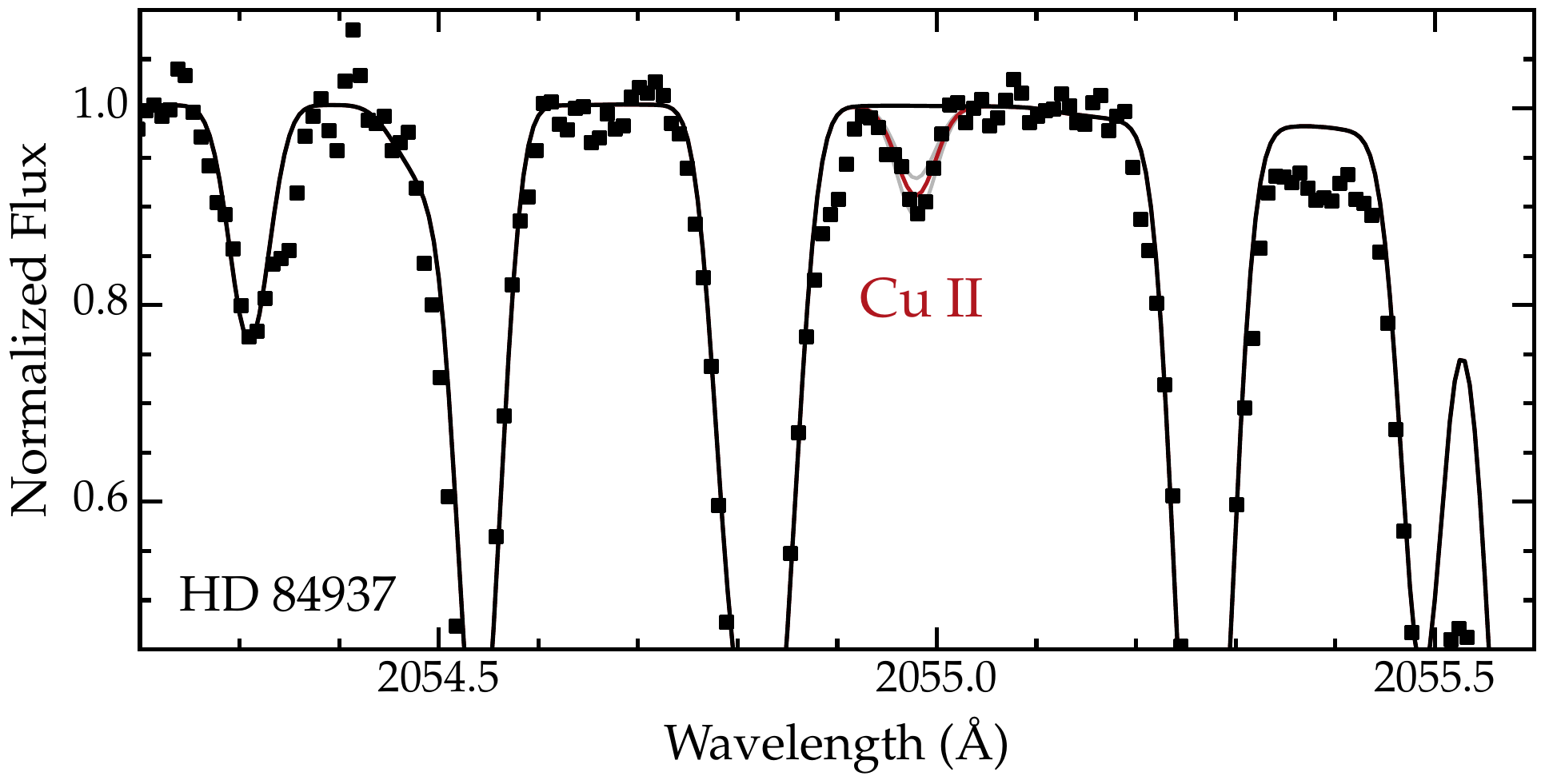} \\
\includegraphics[angle=0,width=3.4in]{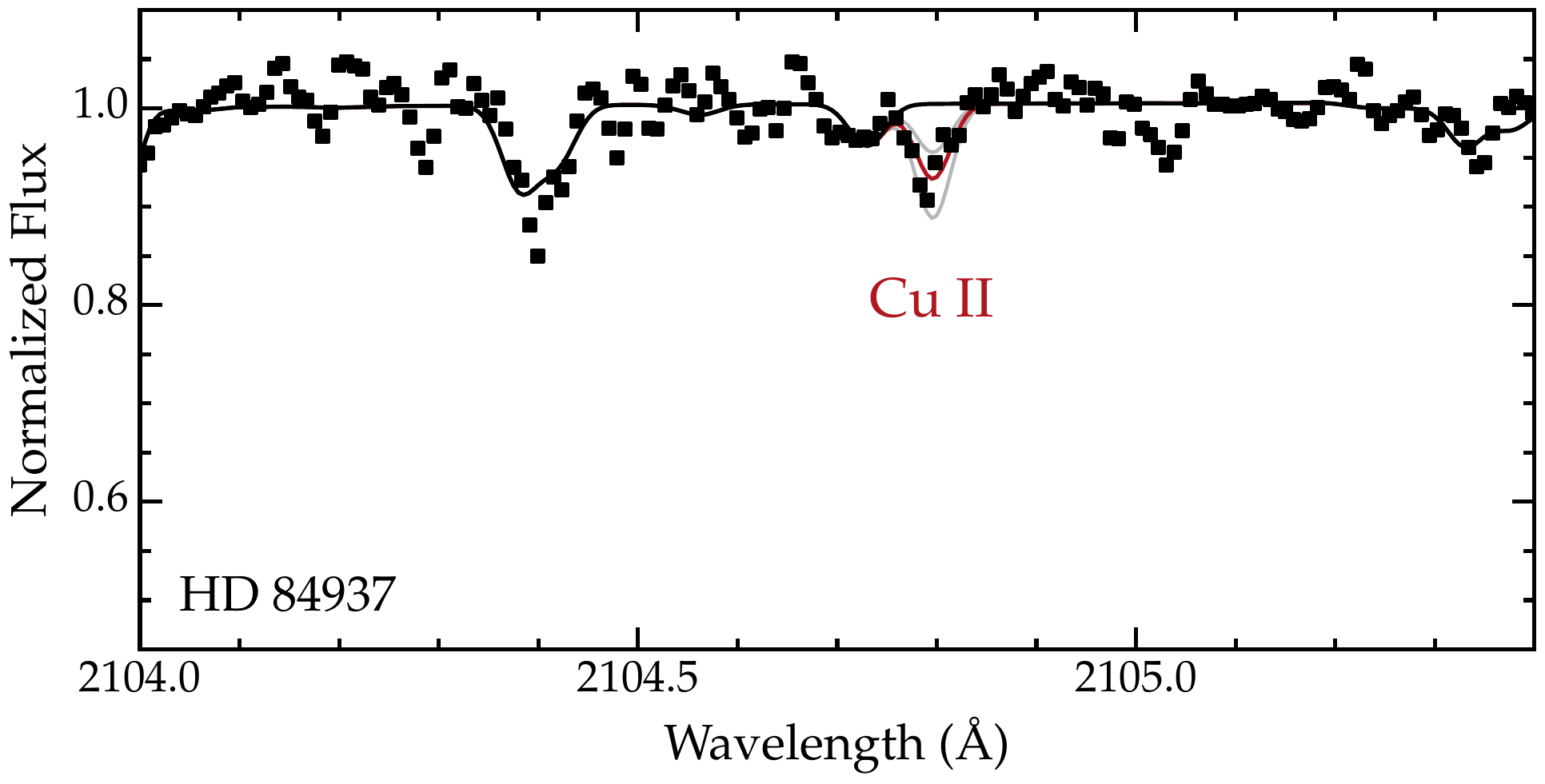} 
\hspace*{0.1in}
\includegraphics[angle=0,width=3.4in]{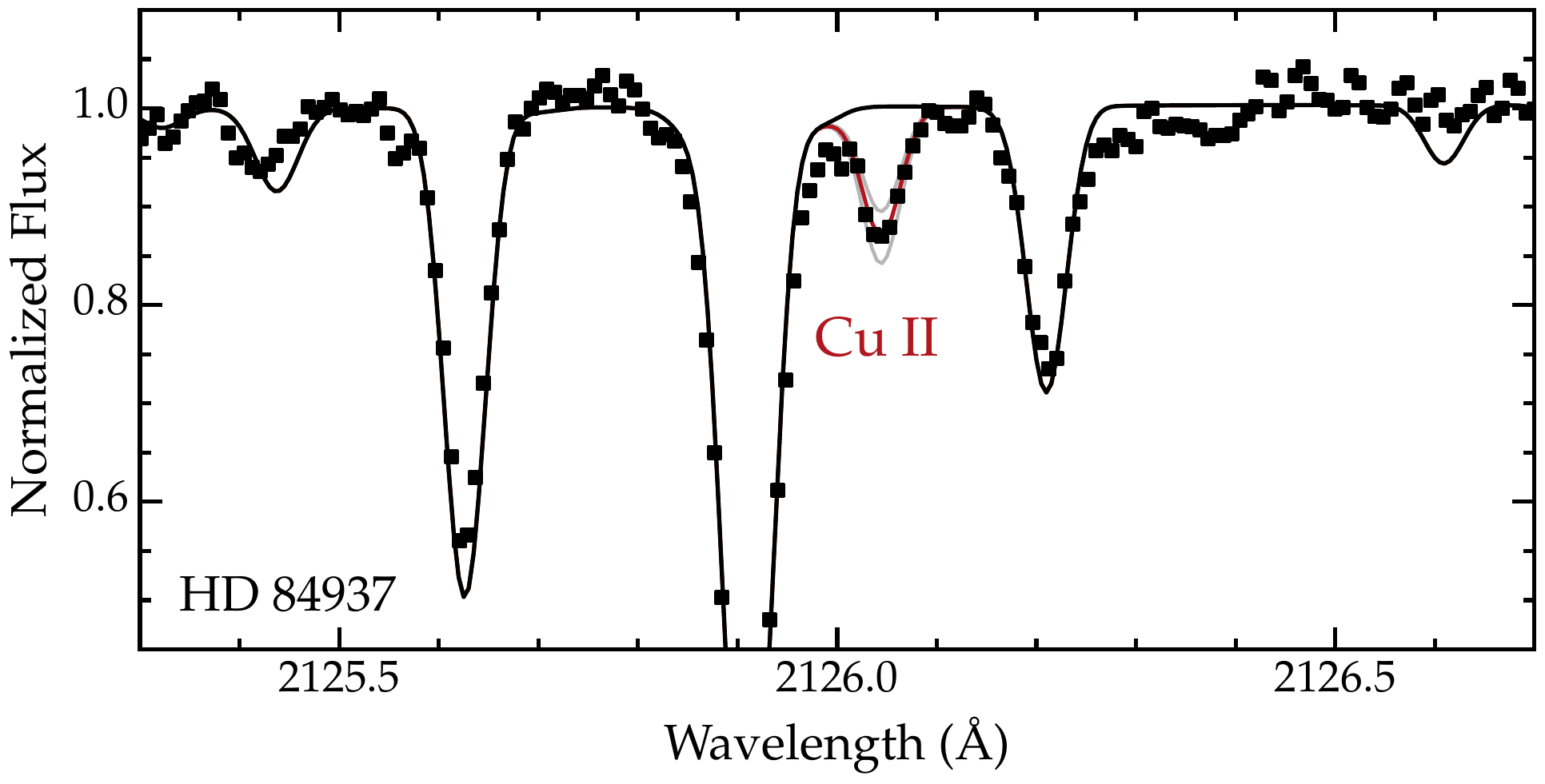} 
\caption{
\label{cuspecplot}
Comparison of observed and synthetic spectra for
Cu~\textsc{i} and \textsc{ii} lines
in \mbox{HD~84937}.
The observed spectra are marked by filled squares.
The red line marks the best-fit model spectrum,
the gray lines mark the $\pm$~1$\sigma$ uncertainties,
and the black line marks a model spectrum with no Cu present.
 }
\end{figure*}

\begin{figure*}
\includegraphics[angle=0,width=3.4in]{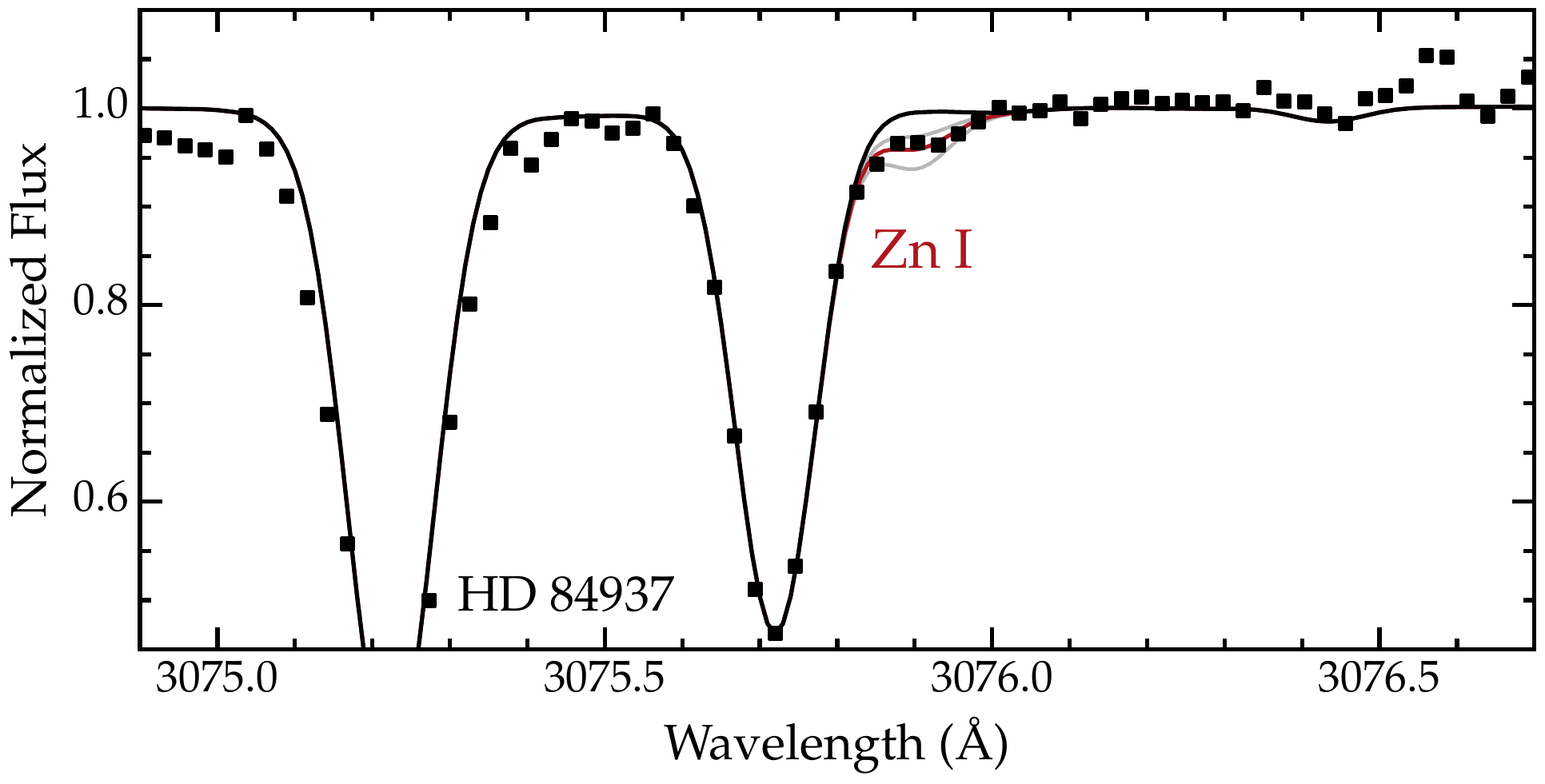} 
\hspace*{0.1in}
\includegraphics[angle=0,width=3.4in]{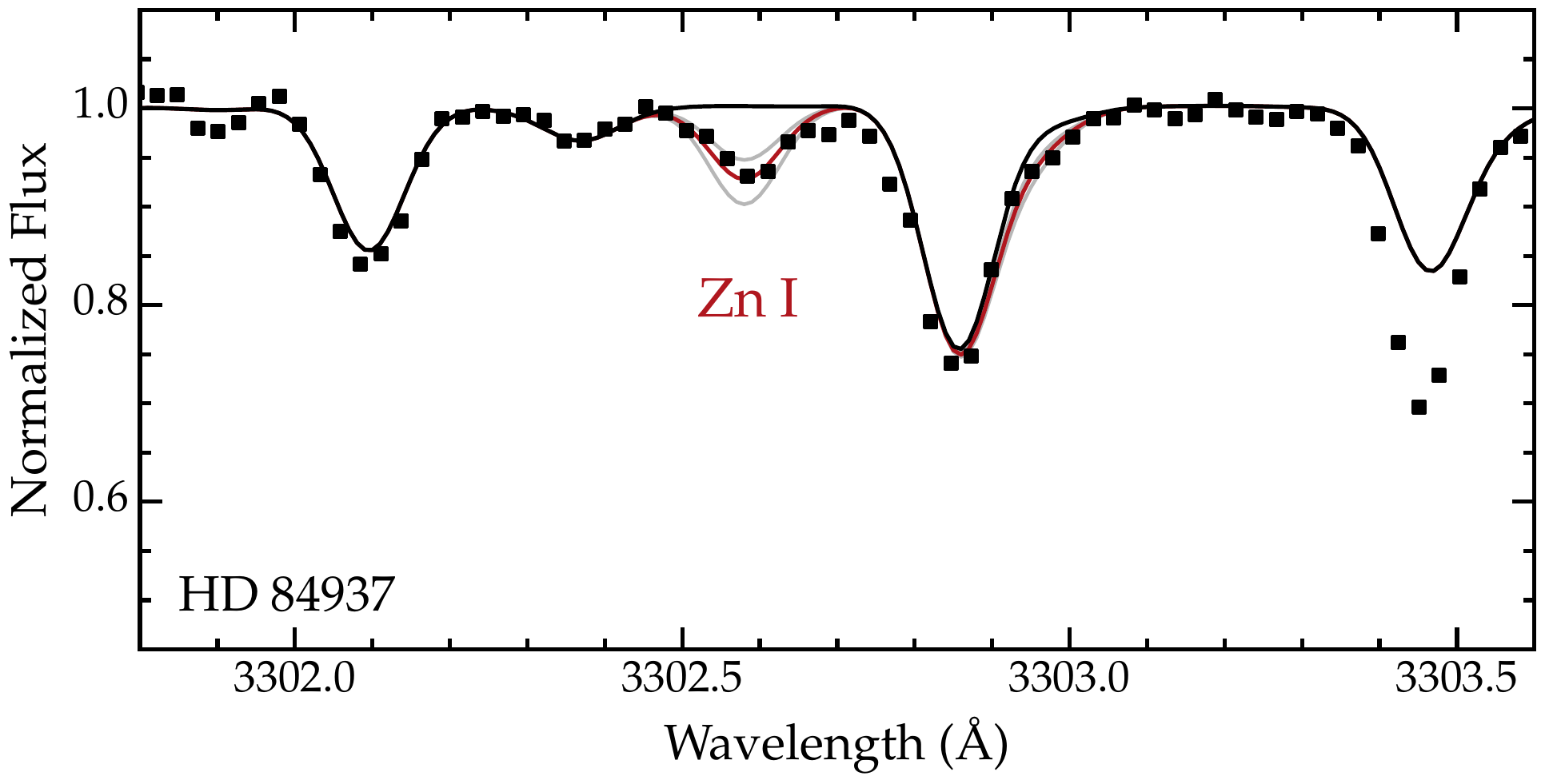} \\
\includegraphics[angle=0,width=3.4in]{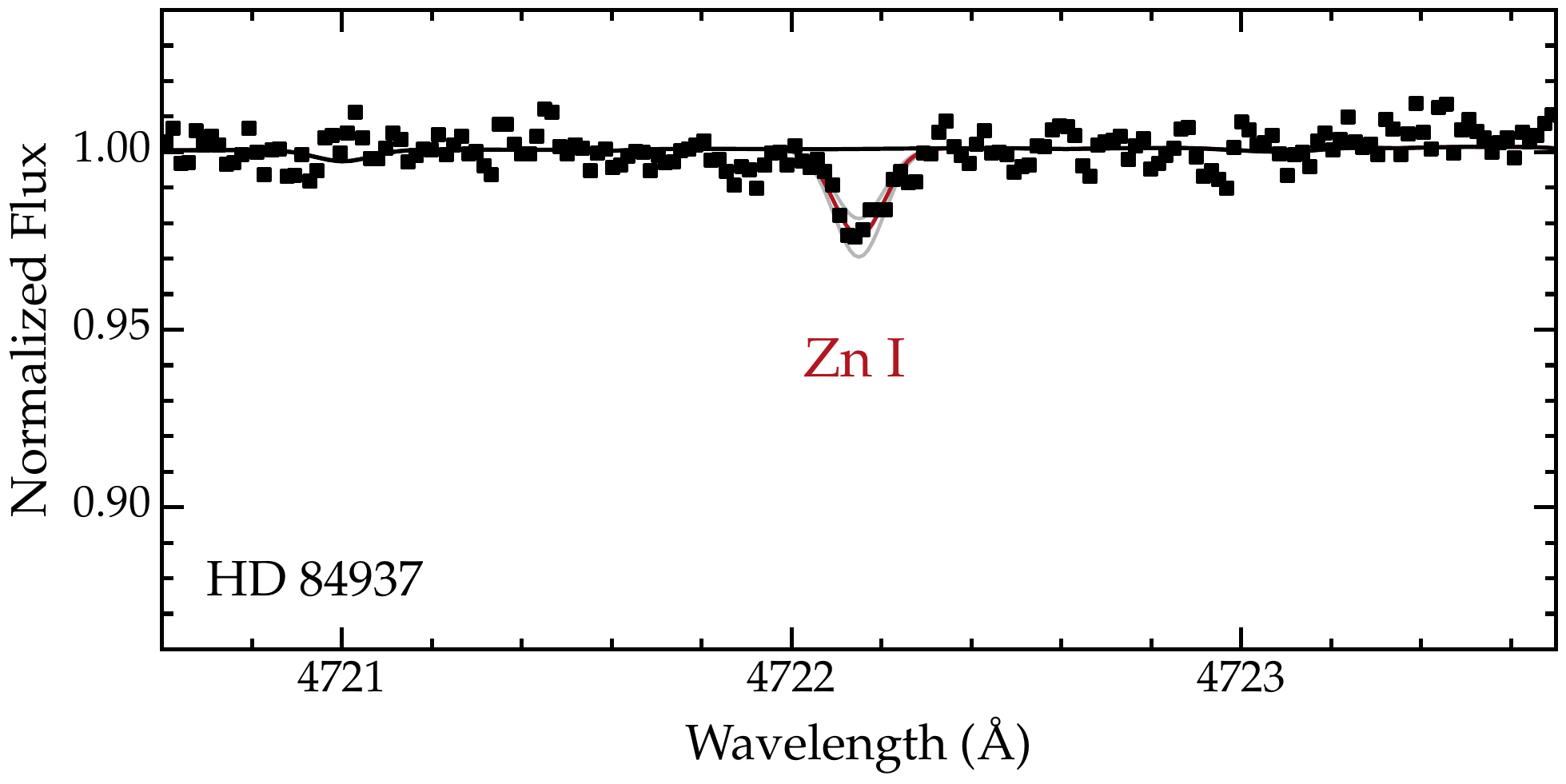} 
\hspace*{0.1in}
\includegraphics[angle=0,width=3.4in]{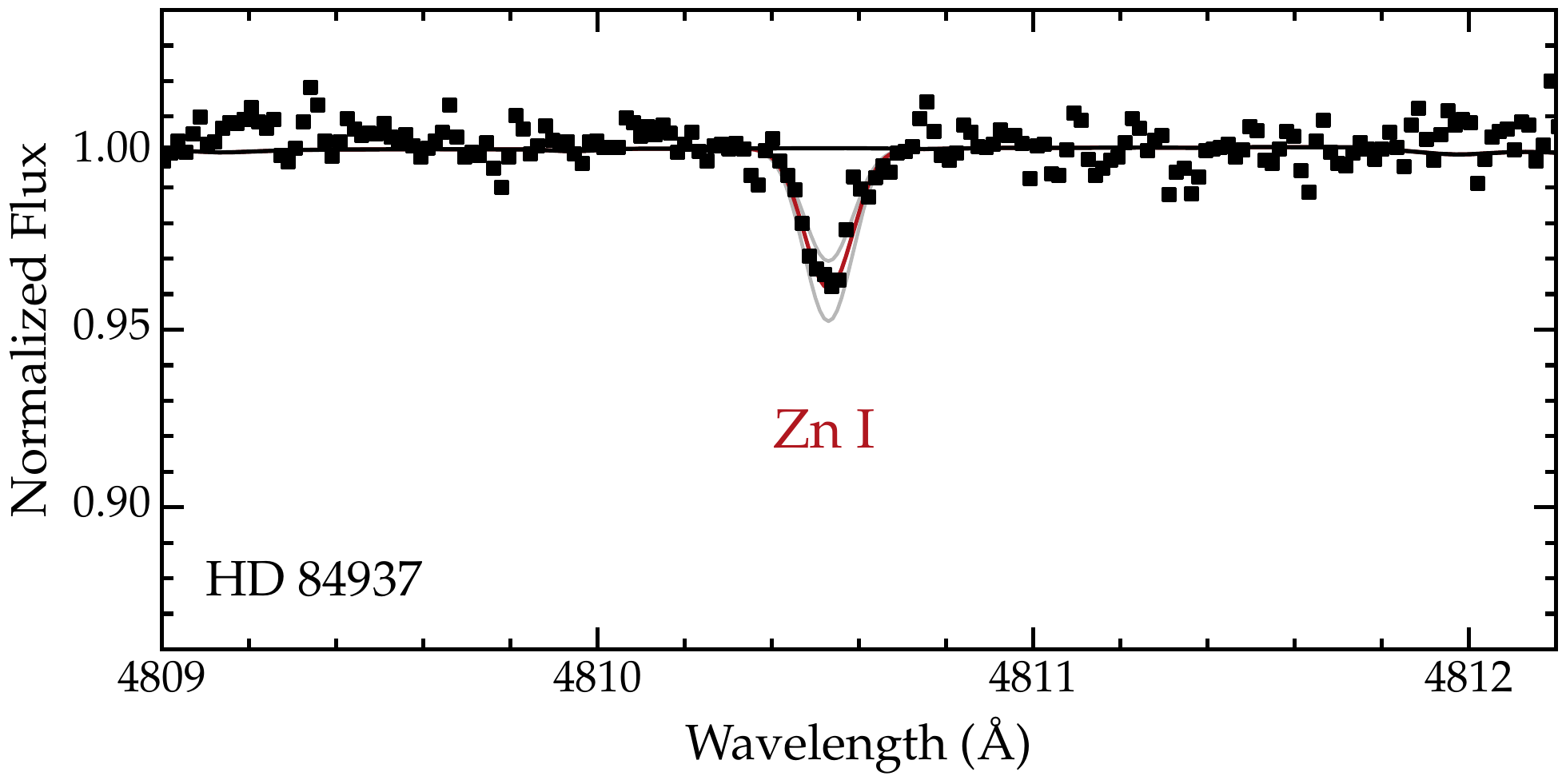} \\
\includegraphics[angle=0,width=3.4in]{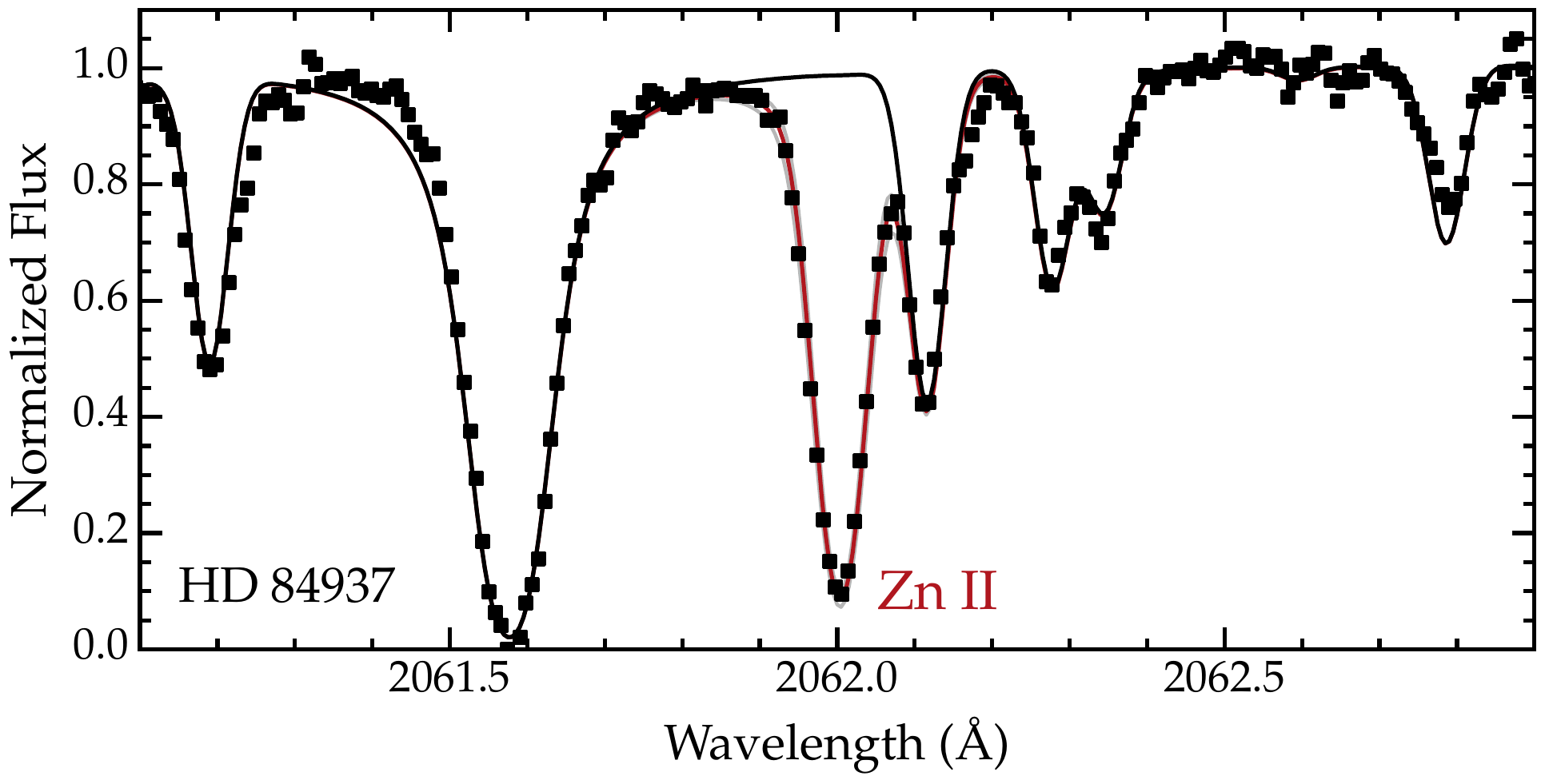} 
\hspace*{0.1in}
\includegraphics[angle=0,width=3.4in]{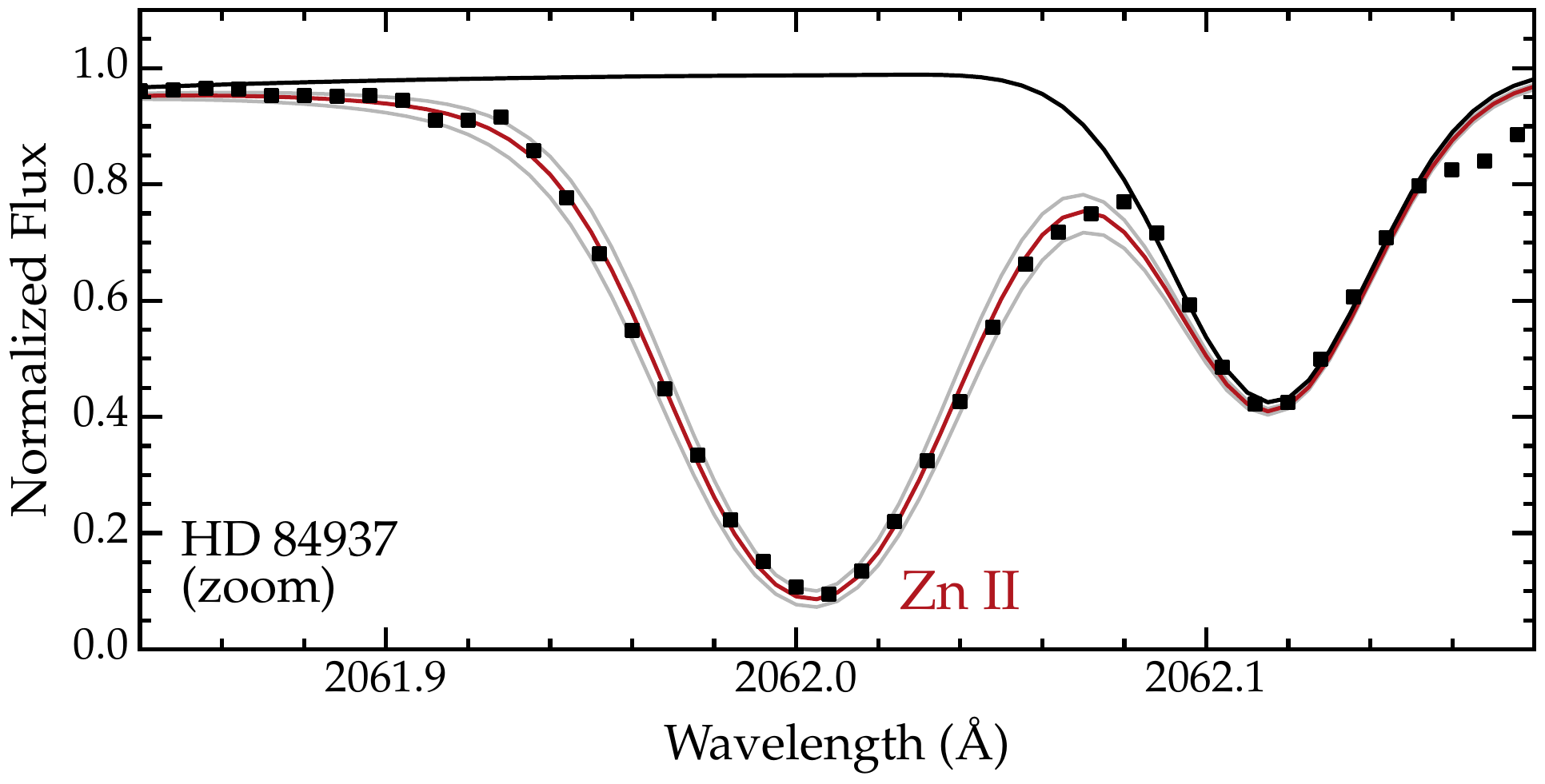} 
\caption{
\label{znspecplot}
Comparison of observed and synthetic spectra for
Zn~\textsc{i} and \textsc{ii} lines
in \mbox{HD~84937}.
The observed spectra are marked by filled squares.
The red line marks the best-fit model spectrum,
the gray lines mark the $\pm$~1$\sigma$ uncertainties,
and the black line marks a model spectrum with no Zn present.
An additional Zn~\textsc{i} line at 3302.94~\AA\ is 
marginally discernible in the upper right panel.
This line is too weak and blended to use 
as an abundance indicator.
The lower right panel shows a zoom-in around the 
Zn~\textsc{ii} line at 2062.00~\AA.
 }
\end{figure*}

Figures~\ref{cuspecplot} and \ref{znspecplot} 
show the Cu and Zn lines used in \hdeightfour.
We do not use the Cu~\textsc{i} lines at 3247 and 3273~\AA\ 
when they saturate
in cooler and more metal-rich stars.
We derive abundances from one or more 
Cu~\textsc{i} resonance lines and
one or more high-excitation lines in only two stars,
but in these two cases we find no significant 
abundance discrepancies.
Commentary on individual lines in the UV spectrum of 
\hdonesixzero\ may be found in 
Section~5 of \citet{roederer12}.
One Zn~\textsc{i} line used in the present study, 
at 2138.56~\AA, was not discussed there.
This line is blended with a Ni~\textsc{ii} line at 2138.58~\AA.~
\citet{fedchak99} report an experimental 
\loggf\ for this Ni~\textsc{ii} line.
The influence of this blending feature is minimal at low metallicity.
Even in higher-metallicity stars, however, 
the high resolving power of the STIS E230H spectra
enables us to derive abundances from
the blue (short) side of the Zn~\textsc{i} line that is largely
unaffected by the blend with Ni~\textsc{ii}.
This line yields abundances in good agreement with
those derived from Zn~\textsc{i} lines in the optical spectrum,
as can be inferred from Table~\ref{zntab}.

\subsection{Comparison with Previous Results}
\label{previousabund}

We now compare our derived [Cu/Fe] and
[Zn/Fe] ratios with the previous studies
listed in Section~\ref{previousparams}.
For [Cu/Fe] in \hdonenine, \hdsevensix, \hdeightfour,
\hdninefour, \hdonefourzero, and \hdonesixzero,
this results in a total of 
2, 3, 1, 2, 1, and 1 derivations in comparison stars.
For [Zn/Fe], the same six stars have a total of 
3, 5, 4, 4, 5, and 2 derivations in comparison stars.
Our derived abundance ratios agree well in most cases,
within 2 standard deviations of the mean of previous analyses.

One exception is [Cu/Fe] in \hdonenine.
We derive [Cu/Fe]~$= -$0.86~$\pm$~0.16 from Cu~\textsc{i} lines,
whereas \citet{mishenina02} derived $-$0.54~$\pm$~0.10.
We use the two Cu~\textsc{i} resonance lines in the blue part of the spectrum,
while \citeauthor{mishenina02}\ used three high-excitation Cu~\textsc{i}
lines in the red part of the spectrum.
None of these high-excitation lines is visible in our HIRES spectrum.
\citeauthor{mishenina02}\ did not report the \loggf\ values adopted,
so we cannot compare these with our own.
The only other study that derived [Cu/Fe] in \hdonenine,
\citet{bihain04}, found
[Cu/Fe]~$= -$0.65~$\pm$~0.14
based on the resonance lines in the blue.
Their result is consistent with both
\citeauthor{mishenina02}\ and us, so
we do not regard this as a serious discrepancy.

The other exception, formally, is [Zn/Fe] in \hdeightfour.
We derive [Zn/Fe]~$= +$0.23~$\pm$~0.14 from Zn~\textsc{i} lines.
Previously,
\citet{sneden91} derived [Zn/Fe]~$= +$0.14~$\pm$~0.10,
\citet{mishenina02} derived [Zn/Fe]~$= +$0.09~$\pm$~0.11,
\citet{nissen07} derived [Zn/Fe]~$= +$0.06~$\pm$~0.06,
and
\citet{bensby14} derived [Zn/Fe]~$= +$0.11~$\pm$~0.27.
We are aware of one additional derivation of 
the [Zn/Fe] ratio in \hdeightfour,
$+$0.16~$\pm$~0.08 \citep{sneden16}.
Our result agrees with four of these five previous studies,
so we are not concerned about the [Zn/Fe] ratio in \hdeightfour.

We noted in Section~\ref{intro} the discrepancy between the
non-LTE calculations of \citet{yan15} and \citet{andrievsky18}
for the Cu~\textsc{i} line at 5105~\AA\ in \hdninefour,
the only line and star common to the two studies.
From their tables and discussion, we infer 
non-LTE $\log\epsilon$(Cu) abundances of
2.37 (\citeauthor{yan15})\ and 2.63 (\citeauthor{andrievsky18})\
for this line.
Our $\log\epsilon$ abundance inferred from the
four Cu~\textsc{ii} lines in \hdninefour\ is
2.34, which favors the \citeauthor{yan15}\ value.
Further evaluation of the two sets of non-LTE calculations
must await the availability of
additional lines in more stars in common.

Four previous studies have also examined Cu~\textsc{ii} 
or Zn~\textsc{ii} lines in one or two stars each per study.
\citet{roederer12} derived
[Cu~\textsc{ii}/Fe] and [Zn~\textsc{ii}/Fe] in \hdonesixzero,
\citet{roederer12c} derived
[Zn~\textsc{ii}/Fe] in \hdninefour\ and \hdonefourzero,
\citet{roederer16ipro} reanalyzed both 
[Cu~\textsc{ii}/Fe] and [Zn~\textsc{ii}/Fe] in \hdninefour,
and
\citet{andrievsky18} derived
[Cu~\textsc{ii}/Fe] in \hdeightfour\ and \hdonefourzero.
Our ratios agree in all cases with those derived by
previous analyses within their mutual 1$\sigma$ uncertainties.
The same STIS spectra were used in all of these studies,
so it is reassuring that there is good agreement.

\section{Discussion}
\label{discussion}

\subsection{Copper}
\label{discusscu}

\begin{figure}
\includegraphics[angle=0,width=3.35in]{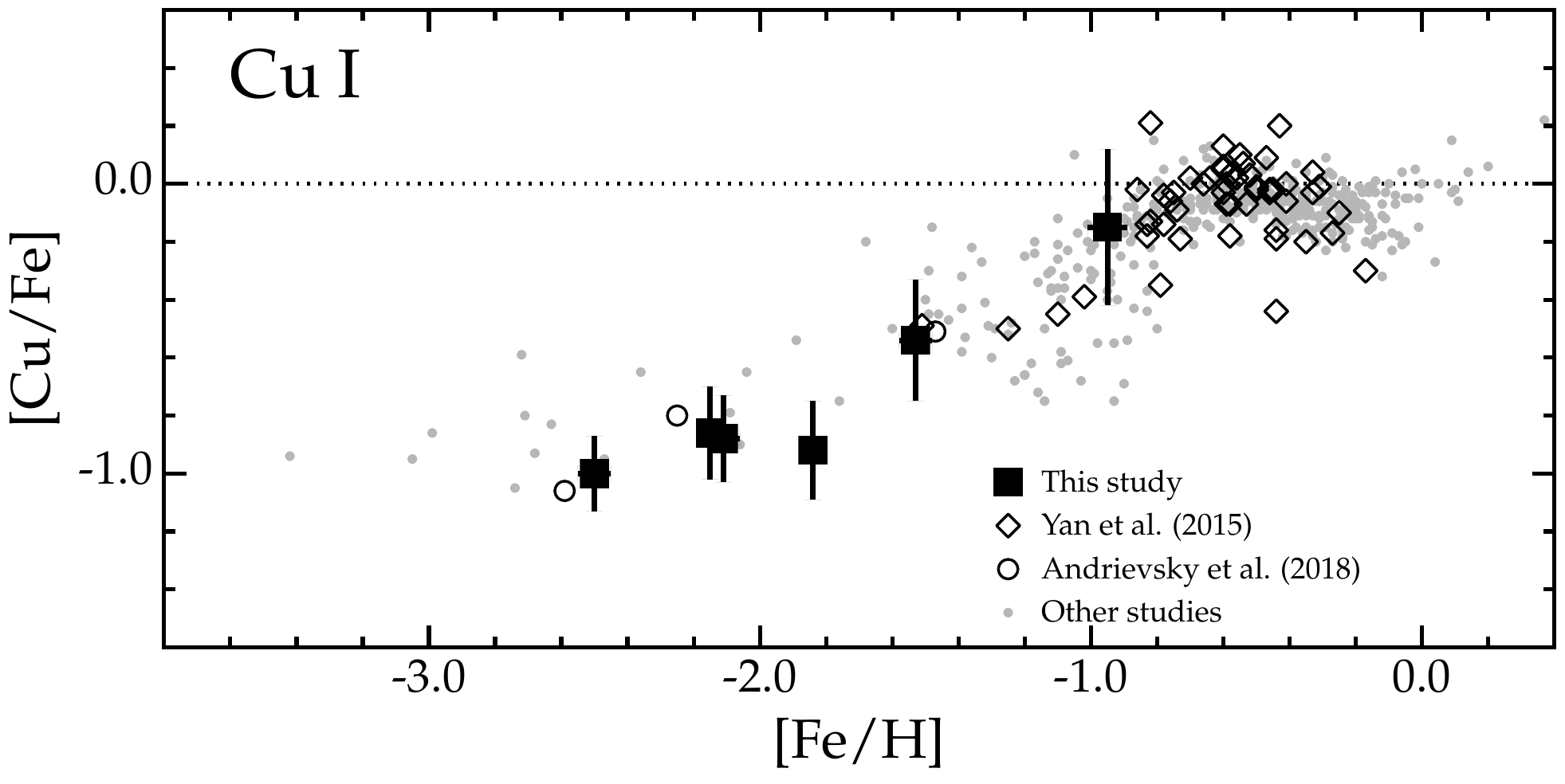} \\
\vspace*{0.1in}
\includegraphics[angle=0,width=3.35in]{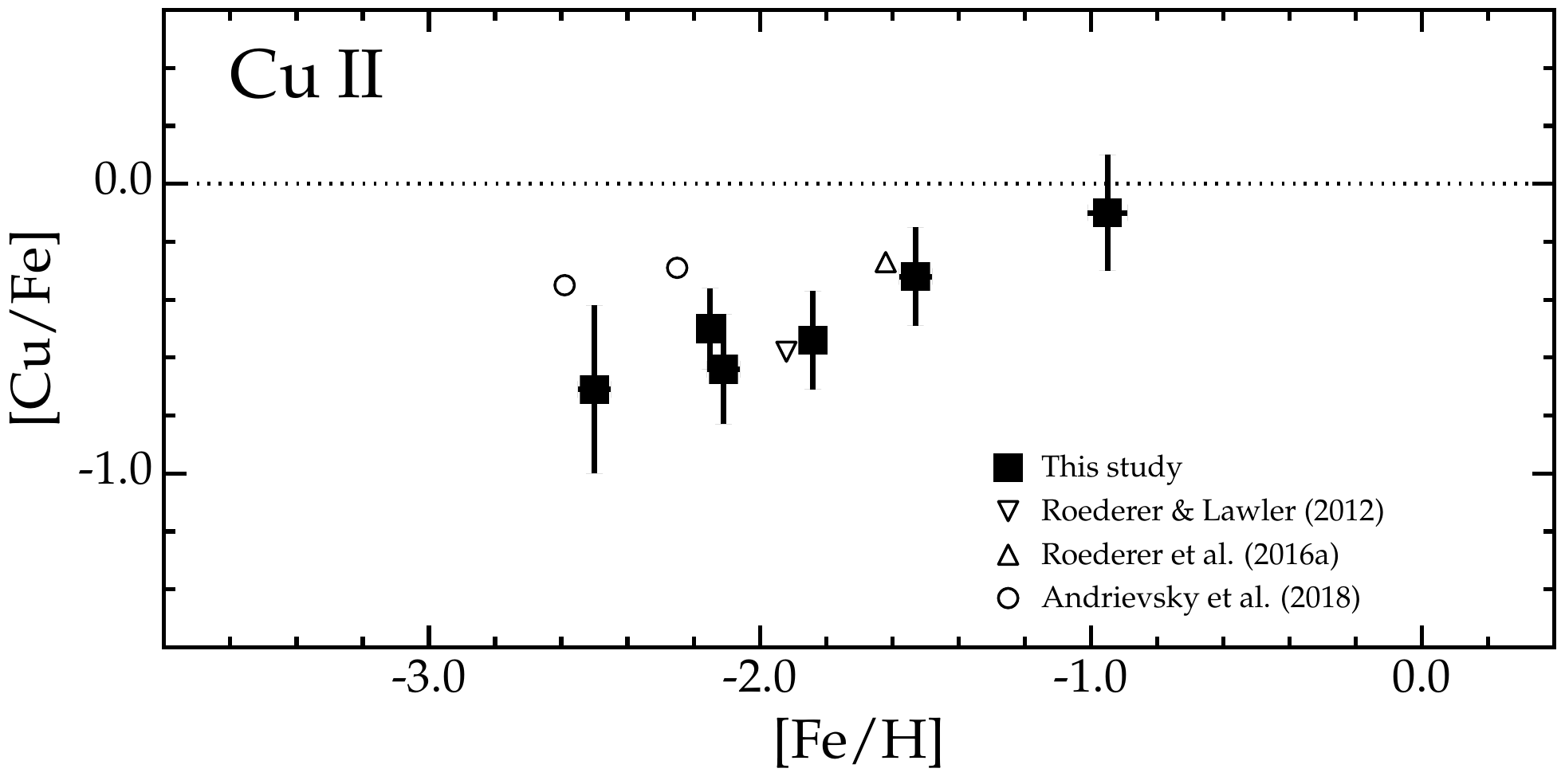} \\
\vspace*{0.1in}
\includegraphics[angle=0,width=3.35in]{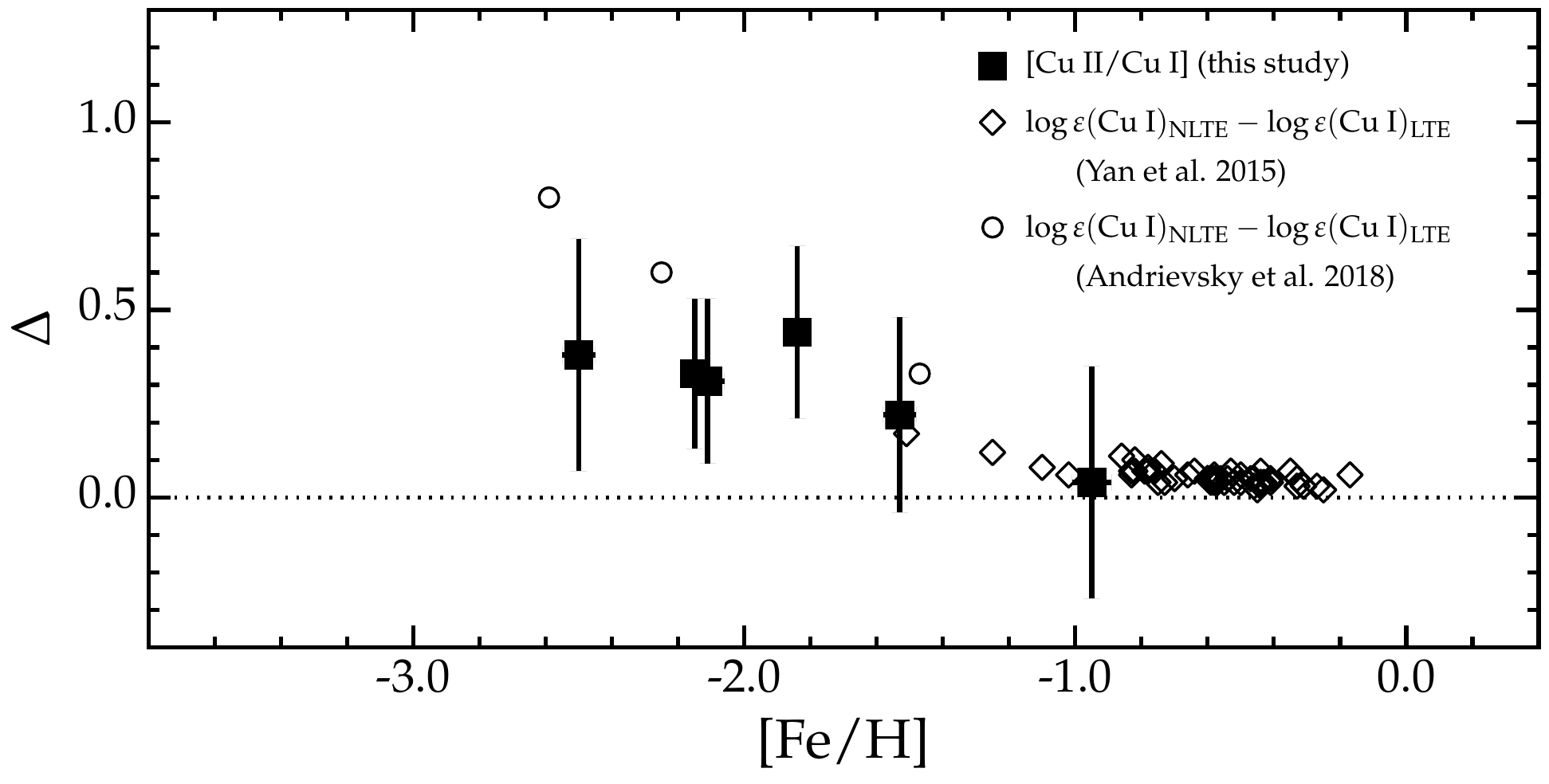} 
\caption{
\label{cuabundplot}
Derived Cu abundances.
The top panel shows [Cu/Fe] derived from Cu~\textsc{i} and Fe~\textsc{i}
lines for the stars in our sample, which are designated 
by filled squares.
The samples of \citet{yan15} and \citet{andrievsky18} 
are designated by open diamonds and circles.
A representative selection of other literature studies of
disk and halo stars is designated by small gray circles.
These data are drawn from 
\citet{sneden91},
\citet{mishenina01,mishenina02},
\citet{reddy03,reddy06},
\citet{bihain04},
\citet{lai08},
\citet{nissen11}, and
\citet{ishigaki13}.
The middle panel shows [Cu/Fe] derived from Cu~\textsc{ii} and Fe~\textsc{ii}
lines for the stars in our sample.
Two stars analyzed by
\citeauthor{andrievsky18}, one star analyzed by \citet{roederer12},
and one star analyzed by \citet{roederer16ipro}
are also shown.
The bottom panel compares our [Cu~\textsc{ii}/Cu~\textsc{i}] ratios
with the Cu~\textsc{i} non-LTE corrections 
presented by \citeauthor{yan15}\ and \citeauthor{andrievsky18} 
Note that the [Cu~\textsc{ii}/Cu~\textsc{i}] ratios
in the bottom panel are calculated from the [Cu/H] ratios, so
they are not a simple difference of the [Cu/Fe] ratios
shown in the top panels.
The dotted line in all panels represents the Solar [Cu/Fe] ratio.
}
\end{figure}

Figure~\ref{cuabundplot} illustrates 
the Cu abundances derived in this study.
The top panel of Figure~\ref{cuabundplot} compares the
[Cu/Fe] ratios derived from neutral lines 
with a representative selection of previous studies from the literature.
The references are listed in the caption to Figure~\ref{cuabundplot}.
We make no attempt to distinguish among stars 
in different Galactic components or to
correct for differences in the atomic data or 
lines used by different authors.
For consistency with the present sample,
only relatively warm (\teff~$>$~5600~K) 
subgiant or main sequence stars are included.
Our derived [Cu/Fe] ratios trace
the trends found in metal-poor stars in the Solar neighborhood,
as revealed by previous studies.
There is a plateau at
[Cu/Fe] ~$= -$0.92~$\pm$~0.04 for the four
stars in our sample with [Fe/H]~$< -$1.8,
a gradual increase of [Cu/Fe] with increasing metallicity,
and another plateau at
[Cu/Fe]~$\approx -$0.1 for comparison stars with [Fe/H]~$> -$0.8.

The middle panel of Figure~\ref{cuabundplot} illustrates the
[Cu/Fe] ratios derived from singly-ionized lines.
The overall trend among stars with [Fe/H]~$\lesssim -$1
is shifted up by a few tenths of a dex
relative to the trend found when using neutral lines.
The [Cu/Fe] ratios remain sub-Solar in the lowest metallicity stars,
and the plateau, if such a description is appropriate,
is found at
[Cu/Fe]~$= -$0.60~$\pm$~0.05 for 
the four stars in our sample with [Fe/H]~$< -$1.8.
The trend of increasing [Cu/Fe] with increasing metallicity
remains, although the magnitude of the slope is reduced.
The [Cu/Fe] ratio in the
most metal-rich star in our sample, \hdsevensix, 
remains consistent with the Solar ratio.

The bottom panel of Figure~\ref{cuabundplot} 
shows two sets of values on the same axes,
and we propose that these two quantities 
probe the same effect.
The black squares show the difference ($\Delta$) between the 
[Cu/H] ratios derived from ionized lines and neutral lines
in our study.
The magnitude of the difference is greatest in the lowest
metallicity stars
($+$0.36~$\pm$~0.06~dex),
and it decreases toward higher metallicities.
We interpret the difference shown in the bottom panel 
as evidence
of the over-ionization of neutral Cu in warm, metal-poor stars,
which leads to an under-prediction of the Cu abundance in LTE.~
The open diamonds and circles
in the bottom panel of Figure~\ref{cuabundplot}
show the theoretical
non-LTE corrections for Cu~\textsc{i} lines calculated
by \citet{yan15} and \citet{andrievsky18}.
The sign and magnitude of the non-LTE corrections
generally match our empirical measurements of the differences between
abundances derived from Cu~\textsc{ii} and Cu~\textsc{i} lines,
although
the \citeauthor{andrievsky18}\ non-LTE calculations
predict slightly larger corrections at the lowest metallicities
studied.
We are encouraged by the overall agreement,
and we propose that these observational signatures generally
validate the theoretical calculations for warm stars by
\citeauthor{yan15}\ and \citeauthor{andrievsky18} 

\subsection{Zinc}
\label{discusszn}

\begin{figure}
\includegraphics[angle=0,width=3.35in]{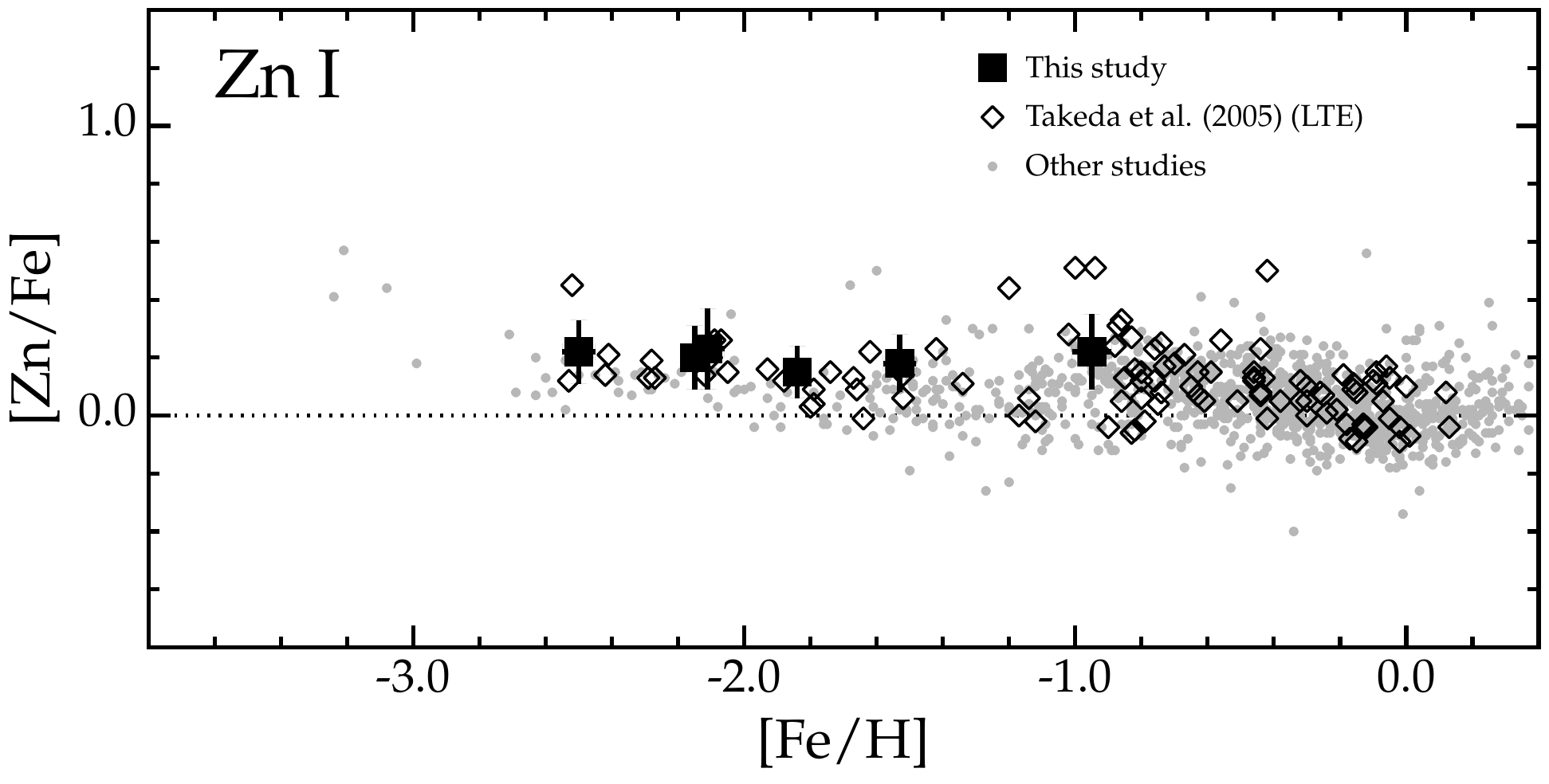} \\
\vspace*{0.1in}
\includegraphics[angle=0,width=3.35in]{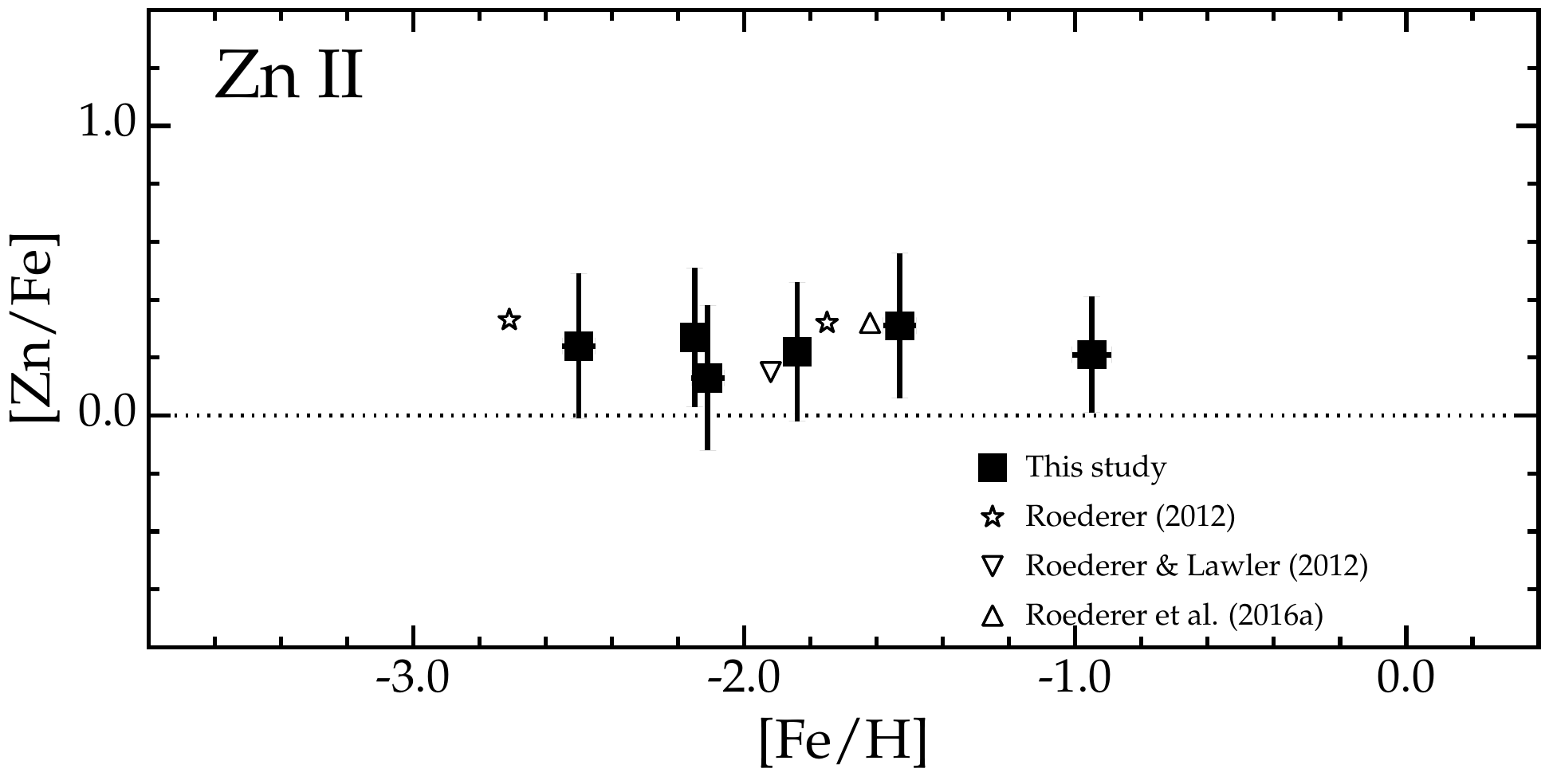} \\
\vspace*{0.1in}
\includegraphics[angle=0,width=3.35in]{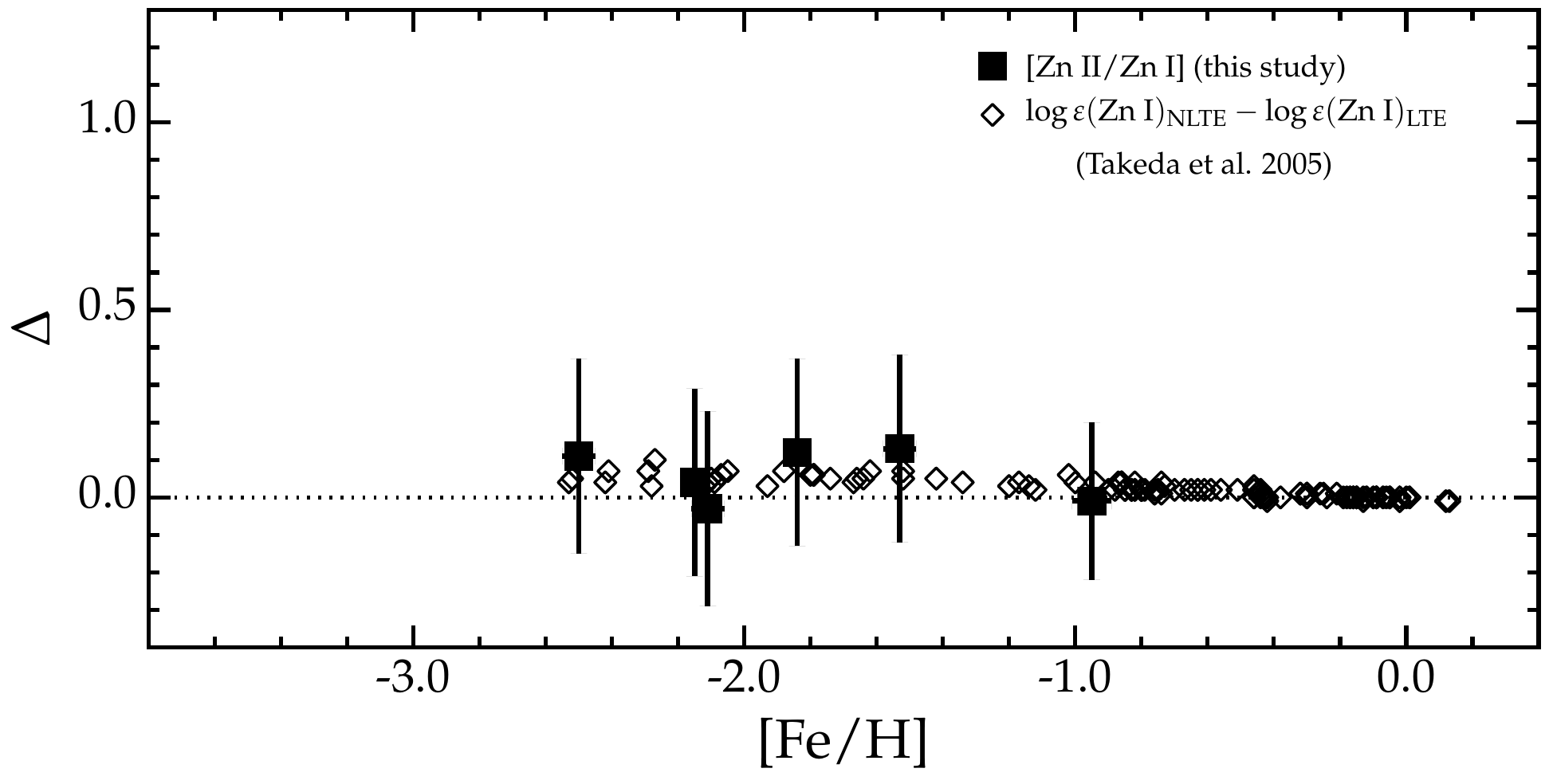} 
\caption{
\label{znabundplot}
Derived Zn abundances.
The top panel shows [Zn/Fe] derived from Zn~\textsc{i} and Fe~\textsc{i}
lines for the stars in our sample, which are designated 
by filled squares.
The sample reanalyzed by \citet{takeda05}
is designated by open diamonds.
A representative selection of other literature studies of
disk and halo stars is designated by small gray circles.
These data are drawn from 
\citet{sneden91},
\citet{mishenina01}, \citet{mishenina02},
\citet{reddy03,reddy06},
\citet{bihain04},
\citet{nissen07},
\citet{lai08},
\citet{nissen11},
\citet{ishigaki13}, and
\citet{bensby14}.
The middle panel shows [Zn/Fe] derived from Zn~\textsc{ii} and Fe~\textsc{ii}
lines for the stars in our sample.
Three stars (re)analyzed by \citet{roederer12c}, \citet{roederer12}, and
\citet{roederer16ipro}
are also shown.
The bottom panel compares our [Zn~\textsc{ii}/Zn~\textsc{i}] ratios
with the Zn~\textsc{i} non-LTE corrections 
presented by \citeauthor{takeda05} 
Note that the [Zn~\textsc{ii}/Zn~\textsc{i}] ratios
in the bottom panel are calculated from the [Zn/H] ratios, so
they are not a simple difference of the [Zn/Fe] ratios
shown in the top panels.
The dotted line in all panels represents the Solar [Zn/Fe] ratio.
}
\end{figure}

Figure~\ref{znabundplot} illustrates the our derived [Zn/Fe] ratios.
A representative selection of [Zn/Fe] ratios
from previous studies of warm, metal-poor stars
is shown for comparison.
The top panel shows the ratios derived from neutral lines.
The six stars in our sample exhibit a consistent enhancement 
of [Zn/Fe]~$= +$0.19~$\pm$~0.02.
As in the case of Cu, the stars in our sample
outline the general trends found in the stars
with similar metallicity in the comparison samples.

The middle panel of Figure~\ref{znabundplot} shows the [Zn/Fe] 
ratios derived from ionized lines. 
The mean [Zn/Fe] ratio is
$+$0.23~$\pm$~0.04.
No trends with metallicity are apparent, 
and the results are in good agreement with previous work.

The black squares in the bottom panel of Figure~\ref{znabundplot}
mark the difference ($\Delta$) between the 
[Zn/H] ratios derived from ionized lines and neutral lines
in our study.
The mildly enhanced [Zn~\textsc{ii}/Zn~\textsc{i}] ratios
found in this study are not significant in any given star.
Considering all six stars as representatives of a single
population, if such a simplification is appropriate,
yields a mean [Zn~\textsc{ii}/Zn~\textsc{i}] ratio of
$+$0.06~$\pm$~0.04~dex,
which is also not significant.
The overall good agreement in the mean
offsets may suggest that our uncertainty estimates
are too conservative.
The open diamonds in the bottom panel of Figure~\ref{znabundplot}
mark Zn~\textsc{i} non-LTE corrections from
\citet{takeda05}.
The non-LTE effects in Zn~\textsc{i} are 
predicted to be small, $<$~0.1~dex, 
and match the differences found by our study.

The agreement between these two quantities is perhaps not surprising.
The first ionization potential of Zn is 9.39~eV,
approximately 1.5~eV higher than the next-highest Fe-group element
(Fe, 7.90~eV; \citealt{morton03}),
and considerably higher than the first ionization potential of Cu
(7.73~eV).
A substantial fraction of Zn atoms in the 
atmospheres of these warm stars is expected to remain neutral
($\approx$~5--50\%; \citealt{roederer12,sneden16}),
in contrast to other Fe-group elements
($\sim$~0.1--5\%; \citeauthor{sneden16}).
Any over-ionization of neutral Zn
has less impact on the derived abundances.
Photon pumping, the bound-bound equivalent of over-ionization,
has been shown to have a notable impact on the level populations
of UV resonance Fe~\textsc{ii} lines \citep{cram80,shchukina05}.
If a similar effect occurs for the Zn~\textsc{ii} UV resonance lines,
it would exacerbate the good agreement in abundances derived from
Zn~\textsc{i} and \textsc{ii} lines.
We conclude that any such photon pumping in the Zn~\textsc{ii}
resonance line at 2062~\AA\ is minimal.

\subsection{HD~84937:\ a Complete Census of Fe-Group Elements
from Majority Species}
\label{hd84937}

\citet{sneden16} derived abundances for Sc through Ni from
1,158 lines of neutral and singly-ionized species
(except for Sc~\textsc{i}, for which no lines were available)
in UV and optical spectra of \hdeightfour,
which is among the UV-brightest metal-poor stars in the sky.
One of the key results of that study
was that LTE abundance calculations 
for most
lines of neutral and ionized species yielded
statistically identical results for all elements
from Ti ($Z =$~22) to Ni ($Z =$~28).
This implies that deviations from
LTE cannot be too large among the Fe-group elements 
in the atmosphere of \hdeightfour\
and presumably other stars with similar stellar parameters
and metallicity.

\begin{figure}
\includegraphics[angle=0,width=3.35in]{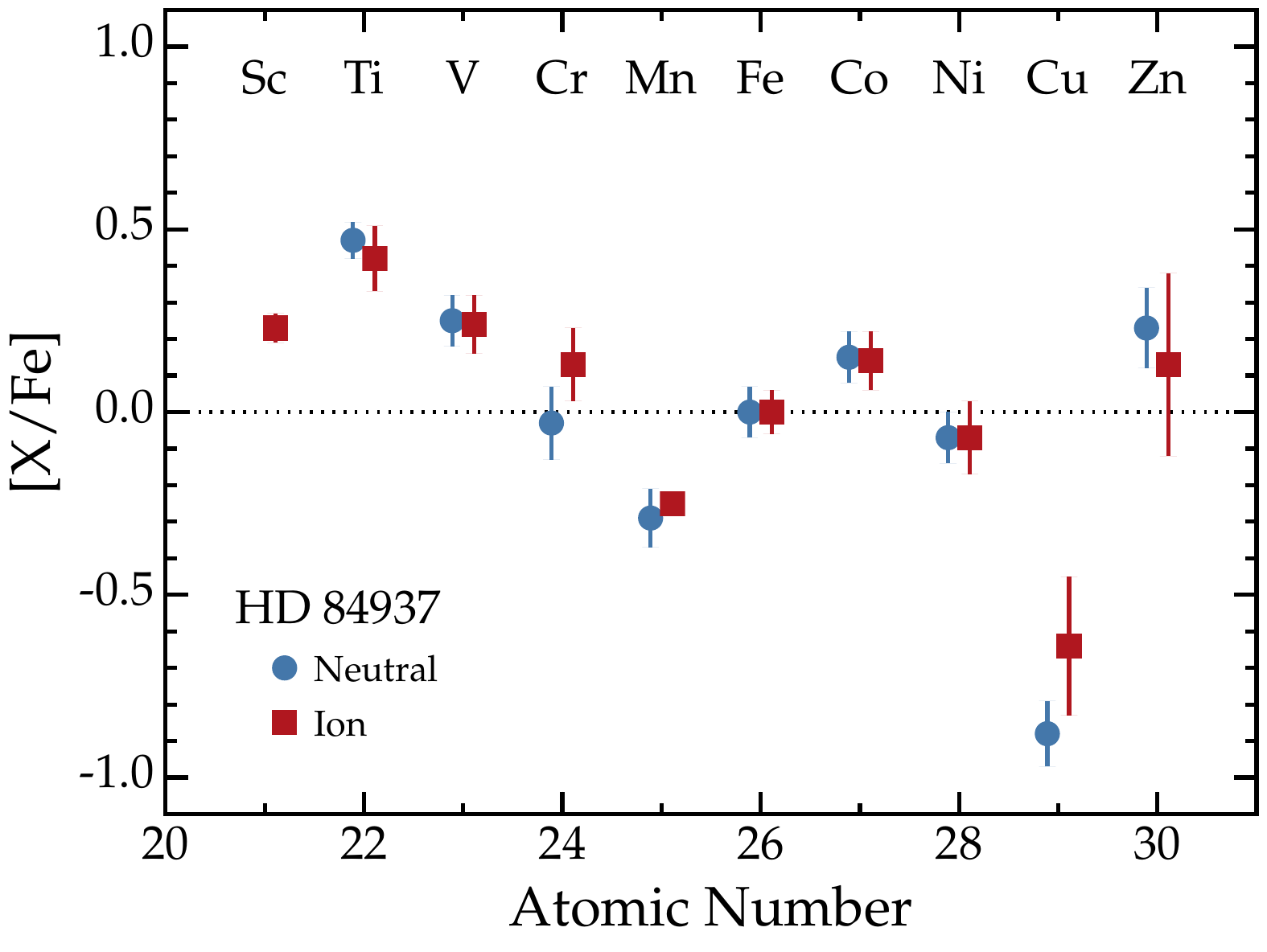} 
\caption{
\label{hd84abundplot}
Comparison of [X/Fe] ratios for elements in the Fe group 
of \mbox{HD~84937}.
The blue circles indicate ratios derived from 
lines of neutral species, and
the red squares indicate ratios derived from 
lines of singly-ionized species.
The data from Sc to Ni are taken from \citet{sneden16},
and the data for Cu and Zn are derived in the present study.
The dotted line marks the Solar ratio.
 }
\end{figure}

\citet{sneden16} also derived [Cu/Fe] and [Zn/Fe] from neutral
species only, because 
no spectrum covering the Cu~\textsc{ii} and Zn~\textsc{ii} lines
was available at the time.
The combined results of \citeauthor{sneden16}\ and our work
are shown in Figure~\ref{hd84abundplot}.
To the best of our knowledge, \hdeightfour\ is the first 
metal-poor star whose complete Fe-group 
(21~$\leq Z \leq$~30) pattern 
has been derived from the majority (i.e., singly ionized) species
of each element.

Our results extend this good agreement to Zn.
The [Zn/Fe] ratio derived from 4~neutral lines is 
determined to a precision of 0.14~dex.
The large abundance uncertainty, 0.25~dex, associated with a
single, saturated UV resonance line of Zn~\textsc{ii}
could mask moderate deviations from LTE.~
There is no evidence, however, for
deviations larger than $\approx$~0.1~dex at present.

Cu is the exception to the good agreement established
by other Fe-group species.
However, the disagreement is only moderately significant,
[Cu~\textsc{ii}/Cu~\textsc{i}] $=$~0.31~$\pm$~0.22~dex,
and it is not apparent in Figure~\ref{hd84abundplot},
where the Fe abundances are folded into the [Cu/Fe] ratios.
The uncertainty is 
dominated by the statistical uncertainty in 
measuring Cu~\textsc{ii} lines,
and it is unlikely that 
significantly higher S/N spectra
can be obtained for \hdeightfour\ or other stars like it 
in the foreseeable future.
The consistent results for Cu~\textsc{i} 
obtained from other metal-poor stars in our sample
suggest that it is likely that LTE calculations
underestimate the abundance derived from Cu~\textsc{i} lines
in \hdeightfour.

\section{Summary}
\label{summary}

We examine archival UV and optical spectra of
six warm (5766~$\leq$~\teff~$\leq$~6427~K),
metal-poor ($-$2.50~$\leq$~[Fe/H]~$\leq -$0.95)
dwarf and subgiant (3.64~$\leq$~\logg~$\leq$~4.44) stars.
We analyze optical and UV
lines from the neutral and singly-ionized species
of Fe, Cu, and Zn to assess the differences that
may result from over-ionization of the
minority neutral species in the atmospheres of these stars.

We conclude that Cu abundances derived from Cu~\textsc{i} lines
in warm, metal-poor stars
may be underestimated by $\approx$~0.36~$\pm$~0.06~dex
at the lowest metallicities, [Fe/H]~$< -$1.8.
At higher metallicities the magnitude of the effect lessens.
Previous theoretical work by
\citet{yan15} and \citet{andrievsky18} 
suggests this underestimation
is caused by the non-LTE over-ionization of neutral Cu.
Our results validate those calculations,
although the \citeauthor{andrievsky18}\ calculations
may slightly overestimate the magnitude 
of the non-LTE corrections at the lowest metallicities.
Additional observational comparisons
would help clarify this matter.

Abundances derived from Zn~\textsc{ii} and Zn~\textsc{i}
differ by only 0.06~$\pm$~0.04~dex on average,
which is not significant.
A substantial fraction of neutral Zn is present, 
unlike in the case of Cu,
minimizing the impact of any over-ionization of Zn.
These results affirm the
non-LTE calculations of \citet{takeda05},
which predict only small non-LTE corrections
for Zn~\textsc{i}.
Thus LTE abundances derived from optical 
Zn~\textsc{i} lines are essentially correct 
to better than 0.1~dex.

Our analysis of Cu~\textsc{ii} and Zn~\textsc{ii} lines
in the metal-poor ([Fe/H]~$= -$2.15) 
turnoff star \hdeightfour\ 
complements the extensive work of \citet{sneden16}
to provide a complete set of 
abundances for all Fe-group elements (Sc through Zn)
derived from the majority singly-ionized species.
The abundances derived from lines of neutral and singly-ionized species
in \hdeightfour\
moderately disagree for only one element, Cu.
\citet{roederer18} have begun to extend
this approach on six additional stars, including
several analyzed in the present study.
We cannot understate the importance of 
access to a high-resolution UV echelle spectrograph
in space for this work.

\acknowledgments

We thank the referee for offering helpful suggestions that have
improved this manuscript.
IUR thanks U.\ Heiter for discussing results in advance of publication.
Generous support for Program AR-15051 has been provided by 
a grant from STScI, which is operated by AURA,
under NASA contract NAS5-26555.
IUR also acknowledges partial support from
grant PHY~14-30152 (Physics Frontier Center/JINA-CEE)
awarded by the U.S.\ National Science Foundation (NSF).~
PSB received financial support from the
Swedish Research Council and the project grant
``The New Milky Way'' from the Knut and Alice Wallenberg Foundation.
This research has made use of NASA's
Astrophysics Data System Bibliographic Services;
the arXiv pre-print server operated by Cornell University;
the SIMBAD and VizieR
databases hosted by the
Strasbourg Astronomical Data Center;
the ASD hosted by NIST;
the MAST at STScI; 
and
IRAF software packages
distributed by the National Optical Astronomy Observatories,
which are operated by AURA,
under cooperative agreement with the NSF.
This work has also made use of data from the European Space Agency (ESA)
mission {\it Gaia} 
 (\url{http://www.cosmos.esa.int/gaia}), 
processed by the {\it Gaia} Data Processing and Analysis Consortium (DPAC,
\url{http://www.cosmos.esa.int/web/gaia/dpac/consortium}). 
Funding for the DPAC has been provided by national institutions, in particular
the institutions participating in the {\it Gaia} Multilateral Agreement.

\facility{ESO:3.6m (HARPS), 
HST (STIS), 
Keck I (HIRES), 
Smith (Tull),
VLT (UVES)}

\software{IRAF \citep{tody93},
matplotlib \citep{hunter07},
MOOG \citep{sneden73},
numpy \citep{vanderwalt11},
R \citep{rsoftware},
scipy \citep{jones01}}

\end{document}

%% file: tab1.tex
\begin{deluxetable*}{lcccccc}
\tablecaption{Characteristics of Archival UV and Optical Spectra
\label{obstab}}
\tablewidth{0pt}
\tabletypesize{\scriptsize}
\tablehead{
\colhead{Star} &
\colhead{Instrument} &
\colhead{Program ID} &
\colhead{Datasets} &
\colhead{PI} &
\colhead{$R$} &
\colhead{S/N\tablenotemark{a}} } 
\startdata
\multicolumn{7}{c}{UV Spectra} \\
\hline
HD~19445      & STIS &  GO-14672& OD65A1010-A8030 & Peterson       & 114,000 & 50 \\
HD~76932      & STIS &  GO-9804 & O8P201010-02040 & Duncan         & 114,000 & 55 \\
HD~84937      & STIS &  GO-14161& OCTKA0010-AD030 & Peterson       & 114,000 & 45 \\
HD~94028      & STIS &  GO-8197 & O5CN01010-03040 & Duncan         & 114,000 & 65 \\
HD~140283     & STIS &  GO-7348 & O55Z01030-01050, 
                                  O55Z02010       & Edvardsson     & 114,000 & 45 \\
HD~160617     & STIS &  GO-8197 & O5CN04010-54050 & Duncan         & 114,000 & 25 \\
\hline\hline
\multicolumn{7}{c}{Optical Spectra} \\
\hline
HD~19445      & HIRES & H283Hr              & \nodata  &  Boesgaard     & 49,000          & 550 \\
HD~76932      & UVES  & 67.D-0439(A)        & \nodata  &  Primas        & 49,600; 57,000\tablenotemark{b}  & 310 \\
HD~84937      & HIRES & C314Hr              & \nodata  &  Gal-Yam       & 49,000          & 190 \\
HD~94028      & Tull  & \tablenotemark{c}   & \nodata  &  Roederer      & 30,000          & 140 \\
HD~140283     & UVES  & 67.D-0439(A)        & \nodata  &  Primas        & 49,600          & \nodata \\
              & HARPS & 080.D-0347(A)       &          &  Heiter        & 115,000         & 230 \\
HD~160617     & UVES  & 65.L-0507(A)        & \nodata  &  Primas        & 49,600          & \nodata \\
              & HIRES & H6aH                &          &  Boesgaard     & 49,000          & 160 \\
\enddata      
\tablenotetext{a}{S/N values are given at 2100~\AA\ and 5000~\AA\
for the UV and optical spectra, respectively}
\tablenotetext{b}{Values for the blue and red components of the spectrum}
\tablenotetext{c}{See \citet{roederer14} for observational details}
\end{deluxetable*}

%% file: tab2-stub.tex
\begin{deluxetable}{lccccc}
\tablecaption{Fe Lines and EWs
\label{fetab}}
\tablewidth{0pt}
\tabletypesize{\scriptsize}
\tablehead{
\colhead{Star} &
\colhead{Species} &
\colhead{$\lambda$} &
\colhead{E.P.} &
\colhead{\loggf} &
\colhead{EW} \\
\colhead{} &
\colhead{} &
\colhead{(\AA)} &
\colhead{(eV)} &
\colhead{} &
\colhead{(m\AA)} }
\startdata
HD19445  &     FeI  &  4001.66  &      2.18  &  $-$1.90  &     12.1 \\ 
HD19445  &     FeI  &  4005.24  &      1.56  &  $-$0.61  &     87.9 \\ 
HD19445  &     FeI  &  4009.71  &      2.22  &  $-$1.25  &     31.0 \\ 
\enddata
\tablecomments{The complete version of Table~\ref{fetab} is available
in the online edition of the journal.
A sample is shown here to illustrate its form and content.}
\end{deluxetable}

%% file: tab3.tex
\begin{deluxetable*}{lcccccccc}
\tablecaption{$V$ Magnitudes, Model Atmosphere Parameters, and Derived Iron Abundances
\label{paramtab}}
\tablewidth{0pt}
\tabletypesize{\scriptsize}
\tablehead{
\colhead{Star} &
\colhead{$V$} &
\colhead{\teff} &
\colhead{\logg} &
\colhead{\vt} &
\colhead{[M/H]} &
\colhead{[Fe~\textsc{i}/H]} &
\colhead{[Fe~\textsc{ii}/H]} &
\colhead{[Fe~\textsc{ii}/H]$-$[Fe~\textsc{i}/H]} \\
\colhead{} &
\colhead{(mag)} &
\colhead{(K)} &
\colhead{[cgs]} &
\colhead{(\kmsec)} &
\colhead{} &
\colhead{} &
\colhead{} &
\colhead{} }
\startdata
HD~19445  & 8.06 & 6070 $\pm$ 76  & 4.44 $\pm$ 0.14 & 1.60 $\pm$ 0.10 & $-$2.2 $\pm$ 0.1 & $-$2.12 $\pm$ 0.05 & $-$2.15 $\pm$ 0.04 & $-$0.03 $\pm$ 0.07 \\
HD~76932  & 5.86 & 5945 $\pm$ 93  & 4.17 $\pm$ 0.11 & 1.10 $\pm$ 0.10 & $-$1.0 $\pm$ 0.1 & $-$0.95 $\pm$ 0.07 & $-$0.95 $\pm$ 0.06 & $+$0.00 $\pm$ 0.10 \\
HD~84937  & 8.32 & 6427 $\pm$ 93  & 4.14 $\pm$ 0.14 & 1.45 $\pm$ 0.10 & $-$2.2 $\pm$ 0.1 & $-$2.16 $\pm$ 0.06 & $-$2.11 $\pm$ 0.05 & $+$0.05 $\pm$ 0.07 \\
HD~94028  & 8.22 & 6097 $\pm$ 74  & 4.34 $\pm$ 0.14 & 1.30 $\pm$ 0.10 & $-$1.6 $\pm$ 0.1 & $-$1.52 $\pm$ 0.05 & $-$1.53 $\pm$ 0.05 & $-$0.01 $\pm$ 0.07 \\
HD~140283 & 7.22 & 5766 $\pm$ 64  & 3.64 $\pm$ 0.13 & 1.30 $\pm$ 0.10 & $-$2.6 $\pm$ 0.1 & $-$2.59 $\pm$ 0.05 & $-$2.50 $\pm$ 0.05 & $+$0.09 $\pm$ 0.06 \\
HD~160617 & 8.74 & 6050 $\pm$ 67  & 3.91 $\pm$ 0.13 & 1.50 $\pm$ 0.10 & $-$1.9 $\pm$ 0.1 & $-$1.89 $\pm$ 0.04 & $-$1.84 $\pm$ 0.04 & $+$0.05 $\pm$ 0.06 \\
\enddata
\end{deluxetable*}

%% file: tab4.tex
\begin{deluxetable}{lccc}
\tablecaption{Atomic Data for Cu and Zn Lines
\label{atomictab}}
\tablewidth{0pt}
\tabletypesize{\scriptsize}
\tablehead{
\colhead{Species} &
\colhead{$\lambda$} &
\colhead{E.P.} &
\colhead{\loggf} \\
\colhead{} &
\colhead{(\AA)} &
\colhead{(eV)} &
\colhead{} }
\startdata
Cu~\textsc{i} & 3247.54 &  0.00 &  $-$0.05 \\
              & 3273.96 &  0.00 &  $-$0.35 \\
              & 5105.54 &  1.39 &  $-$1.50 \\
              & 5218.20 &  3.82 &  $+$0.26 \\
Cu~\textsc{ii}& 2037.13 &  2.83 &  $-$0.23 \\
              & 2054.98 &  2.83 &  $-$0.29 \\
              & 2104.80 &  2.98 &  $-$0.51 \\
              & 2126.04 &  2.83 &  $-$0.23 \\
Zn~\textsc{i} & 2138.56 &  0.00 &  $+$0.16 \\
              & 3075.90 &  0.00 &  $-$3.85 \\
              & 3302.58 &  4.03 &  $-$0.02 \\
              & 4680.14 &  4.01 &  $-$0.85 \\
              & 4722.15 &  4.03 &  $-$0.37 \\
              & 4810.53 &  4.08 &  $-$0.15 \\
Zn~\textsc{ii}& 2062.00 &  0.00 &  $-$0.29 \\
\enddata
\end{deluxetable}

%% file: tab5.tex
\begin{deluxetable*}{lccccccccc}
\tablecaption{Derived $\log \epsilon$ Abundances
and Uncertainties for Individual Cu Lines
\label{cutab}}
\tablewidth{0pt}
\tabletypesize{\scriptsize}
\tablehead{
\colhead{Star} &
\multicolumn{4}{c}{Cu~\textsc{i}} &
\colhead{} &
\multicolumn{4}{c}{Cu~\textsc{ii}} \\
\cline{2-5} 
\cline{7-10}
\colhead{} &
\colhead{3247.54~\AA} &
\colhead{3273.96~\AA} &
\colhead{5105.54~\AA} &
\colhead{5218.20~\AA} &
\colhead{} &
\colhead{2037.13~\AA} &
\colhead{2054.98~\AA} &
\colhead{2104.80~\AA} &
\colhead{2126.04~\AA} }
\startdata
HD~19445  & 1.17 $\pm$ 0.10 & 1.25 $\pm$ 0.10 & \nodata         & \nodata         & & 1.51 $\pm$ 0.10 & 1.55 $\pm$ 0.15 & 1.59 $\pm$ 0.15 & 1.57 $\pm$ 0.15 \\
HD~76932  & \nodata         & 3.07 $\pm$ 0.20 & 3.07 $\pm$ 0.05 & 3.15 $\pm$ 0.10 & & 3.13 $\pm$ 0.20 & 3.07 $\pm$ 0.20 & 3.17 $\pm$ 0.20 & 3.19 $\pm$ 0.25 \\
HD~84937  & 1.15 $\pm$ 0.10 & 1.13 $\pm$ 0.05 & \nodata         & \nodata         & & 1.37 $\pm$ 0.20 & 1.39 $\pm$ 0.10 & 1.57 $\pm$ 0.20 & 1.47 $\pm$ 0.10 \\
HD~94028  & 2.13 $\pm$ 0.25 & 2.09 $\pm$ 0.20 & 2.17 $\pm$ 0.20 & \nodata         & & 2.27 $\pm$ 0.15 & 2.33 $\pm$ 0.15 & 2.45 $\pm$ 0.15 & 2.37 $\pm$ 0.15 \\
HD~140283 & 0.57 $\pm$ 0.10 & 0.61 $\pm$ 0.05 & \nodata         & \nodata         & & \nodata         & 0.99 $\pm$ 0.20 & \nodata         & 0.97 $\pm$ 0.10 \\
HD~160617 & 1.31 $\pm$ 0.15 & 1.41 $\pm$ 0.10 & \nodata         & \nodata         & & 1.75 $\pm$ 0.15 & 1.81 $\pm$ 0.15 & 1.92 $\pm$ 0.15 & 1.83 $\pm$ 0.15 \\
\enddata
\tablecomments{The stated uncertainties reflect the quality of the fit 
to the line profile, including factors like noise,
blending features, and continuum placement.}
\end{deluxetable*}

%% file: tab6.tex
\begin{deluxetable*}{lcccccccc}
\tablecaption{Derived $\log \epsilon$ Abundances 
and Uncertainties for Individual Zn Lines
\label{zntab}}
\tablewidth{0pt}
\tabletypesize{\scriptsize}
\tablehead{
\colhead{Star} &
\multicolumn{6}{c}{Zn~\textsc{i}} &
\colhead{} &
\colhead{Zn~\textsc{ii}} \\
\cline{2-7} 
\cline{9-9}
\colhead{} &
\colhead{2138.56~\AA} &
\colhead{3075.90~\AA} &
\colhead{3302.58~\AA} &
\colhead{4680.14~\AA} &
\colhead{4722.15~\AA} &
\colhead{4810.53~\AA} &
\colhead{} &
\colhead{2062.00~\AA} }
\startdata
HD~19445  & 2.68 $\pm$ 0.20 & 2.84 $\pm$ 0.25 & 2.58 $\pm$ 0.20 & 2.68 $\pm$ 0.15 & 2.64 $\pm$ 0.10 & 2.62 $\pm$ 0.05 & & 2.68 $\pm$ 0.15 \\
HD~76932  & 3.80 $\pm$ 0.25 & \nodata         & 3.64 $\pm$ 0.20 & 3.80 $\pm$ 0.10 & 3.84 $\pm$ 0.05 & 3.84 $\pm$ 0.05 & & 3.82 $\pm$ 0.10 \\
HD~84937  & \nodata         & 2.74 $\pm$ 0.20 & 2.82 $\pm$ 0.15 & \nodata         & 2.52 $\pm$ 0.10 & 2.56 $\pm$ 0.10 & & 2.58 $\pm$ 0.15 \\
HD~94028  & 3.30 $\pm$ 0.20 & \nodata         & 3.22 $\pm$ 0.15 & 3.20 $\pm$ 0.10 & 3.20 $\pm$ 0.05 & 3.22 $\pm$ 0.05 & & 3.34 $\pm$ 0.15 \\
HD~140283 & 2.18 $\pm$ 0.20 & 2.30 $\pm$ 0.15 & 2.14 $\pm$ 0.15 & 2.08 $\pm$ 0.20 & 2.24 $\pm$ 0.05 & 2.16 $\pm$ 0.05 & & 2.30 $\pm$ 0.15 \\
HD~160617 & 2.99 $\pm$ 0.15 & 2.93 $\pm$ 0.10 & 2.82 $\pm$ 0.10 & 2.73 $\pm$ 0.20 & 2.82 $\pm$ 0.05 & 2.79 $\pm$ 0.05 & & 2.94 $\pm$ 0.15 \\
\enddata
\tablecomments{The stated uncertainties reflect the quality of the fit 
to the line profile, including factors like noise,
blending features, and continuum placement.}
\end{deluxetable*}

%% file: tab7.tex
\begin{deluxetable*}{lccccc}
\tablecaption{Derived Cu Abundance Ratios
\label{cuabundtab}}
\tablewidth{0pt}
\tabletypesize{\scriptsize}
\tablehead{
\colhead{Star} &
\colhead{[Cu~\textsc{i}/H]} &
\colhead{[Cu~\textsc{i}/Fe~\textsc{i}]} &
\colhead{[Cu~\textsc{ii}/H]} &
\colhead{[Cu~\textsc{ii}/Fe~\textsc{ii}]} &
\colhead{[Cu~\textsc{ii}/H]$-$[Cu~\textsc{i}/H]} \\
\colhead{} &
\colhead{$\pm$ stat.\ $\pm$ sys.} &
\colhead{$\pm$ stat.\ $\pm$ sys.} &
\colhead{$\pm$ stat.\ $\pm$ sys.} &
\colhead{$\pm$ stat.\ $\pm$ sys.} &
\colhead{$\pm$ stat.\ $\pm$ sys.} 
}
\startdata
HD~19445  & $-$2.98 $\pm$ 0.09 $\pm$ 0.11 & $-$0.86 $\pm$ 0.10 $\pm$ 0.12 & $-$2.65 $\pm$ 0.14 $\pm$ 0.05 & $-$0.50 $\pm$ 0.14 $\pm$ 0.06 & $+$0.33 $\pm$ 0.16 $\pm$ 0.12 \\
HD~76932  & $-$1.10 $\pm$ 0.23 $\pm$ 0.10 & $-$0.15 $\pm$ 0.24 $\pm$ 0.12 & $-$1.05 $\pm$ 0.19 $\pm$ 0.07 & $-$0.10 $\pm$ 0.20 $\pm$ 0.09 & $+$0.04 $\pm$ 0.29 $\pm$ 0.12 \\
HD~84937  & $-$3.06 $\pm$ 0.09 $\pm$ 0.10 & $-$0.88 $\pm$ 0.09 $\pm$ 0.12 & $-$2.75 $\pm$ 0.18 $\pm$ 0.05 & $-$0.64 $\pm$ 0.19 $\pm$ 0.07 & $+$0.31 $\pm$ 0.19 $\pm$ 0.11 \\
HD~94028  & $-$2.07 $\pm$ 0.16 $\pm$ 0.11 & $-$0.54 $\pm$ 0.17 $\pm$ 0.12 & $-$1.85 $\pm$ 0.16 $\pm$ 0.05 & $-$0.32 $\pm$ 0.17 $\pm$ 0.07 & $+$0.22 $\pm$ 0.23 $\pm$ 0.12 \\
HD~140283 & $-$3.59 $\pm$ 0.06 $\pm$ 0.09 & $-$1.00 $\pm$ 0.08 $\pm$ 0.10 & $-$3.21 $\pm$ 0.28 $\pm$ 0.05 & $-$0.71 $\pm$ 0.29 $\pm$ 0.07 & $+$0.38 $\pm$ 0.29 $\pm$ 0.10 \\
HD~160617 & $-$2.81 $\pm$ 0.10 $\pm$ 0.12 & $-$0.92 $\pm$ 0.11 $\pm$ 0.13 & $-$2.38 $\pm$ 0.16 $\pm$ 0.04 & $-$0.54 $\pm$ 0.17 $\pm$ 0.06 & $+$0.44 $\pm$ 0.19 $\pm$ 0.13 \\
\enddata
\tablecomments{The statistical uncertainties (``stat.'')\ include contributions
from line fitting and \loggf\ values.
The systematic uncertainties (``sys.'')\ include contributions
from uncertainties in the model atmosphere parameters.}
\end{deluxetable*}

%% file: tab8.tex
\begin{deluxetable*}{lccccc}
\tablecaption{Derived Zn Abundance Ratios
\label{znabundtab}}
\tablewidth{0pt}
\tabletypesize{\scriptsize}
\tablehead{
\colhead{Star} &
\colhead{[Zn~\textsc{i}/H]} &
\colhead{[Zn~\textsc{i}/Fe~\textsc{i}]} &
\colhead{[Zn~\textsc{ii}/H]} &
\colhead{[Zn~\textsc{ii}/Fe~\textsc{ii}]} &
\colhead{[Zn~\textsc{ii}/H]$-$[Zn~\textsc{i}/H]} \\
\colhead{} &
\colhead{$\pm$ stat.\ $\pm$ sys.} &
\colhead{$\pm$ stat.\ $\pm$ sys.} &
\colhead{$\pm$ stat.\ $\pm$ sys.} &
\colhead{$\pm$ stat.\ $\pm$ sys.} &
\colhead{$\pm$ stat.\ $\pm$ sys.} 
}
\startdata
HD~19445  & $-$1.92 $\pm$ 0.07 $\pm$ 0.05 & $+$0.20 $\pm$ 0.09 $\pm$ 0.07 & $-$1.88 $\pm$ 0.24 $\pm$ 0.06 & $+$0.27 $\pm$ 0.24 $\pm$ 0.07 & $+$0.04 $\pm$ 0.25 $\pm$ 0.08 \\
HD~76932  & $-$0.73 $\pm$ 0.06 $\pm$ 0.06 & $+$0.22 $\pm$ 0.09 $\pm$ 0.09 & $-$0.74 $\pm$ 0.20 $\pm$ 0.07 & $+$0.21 $\pm$ 0.20 $\pm$ 0.09 & $-$0.01 $\pm$ 0.21 $\pm$ 0.09 \\
HD~84937  & $-$1.95 $\pm$ 0.09 $\pm$ 0.06 & $+$0.23 $\pm$ 0.11 $\pm$ 0.09 & $-$1.98 $\pm$ 0.24 $\pm$ 0.09 & $+$0.13 $\pm$ 0.25 $\pm$ 0.10 & $-$0.03 $\pm$ 0.26 $\pm$ 0.11 \\
HD~94028  & $-$1.35 $\pm$ 0.06 $\pm$ 0.04 & $+$0.18 $\pm$ 0.08 $\pm$ 0.06 & $-$1.22 $\pm$ 0.24 $\pm$ 0.06 & $+$0.31 $\pm$ 0.25 $\pm$ 0.08 & $+$0.13 $\pm$ 0.25 $\pm$ 0.07 \\
HD~140283 & $-$2.37 $\pm$ 0.06 $\pm$ 0.05 & $+$0.22 $\pm$ 0.08 $\pm$ 0.07 & $-$2.26 $\pm$ 0.24 $\pm$ 0.08 & $+$0.24 $\pm$ 0.25 $\pm$ 0.09 & $+$0.11 $\pm$ 0.26 $\pm$ 0.09 \\
HD~160617 & $-$1.74 $\pm$ 0.06 $\pm$ 0.05 & $+$0.15 $\pm$ 0.07 $\pm$ 0.06 & $-$1.62 $\pm$ 0.24 $\pm$ 0.07 & $+$0.22 $\pm$ 0.24 $\pm$ 0.08 & $+$0.12 $\pm$ 0.25 $\pm$ 0.09 \\
\enddata
\tablecomments{The statistical uncertainties (``stat.'')\ include contributions
from line fitting and \loggf\ values.
The systematic uncertainties (``sys.'')\ include contributions
from uncertainties in the model atmosphere parameters.}
\end{deluxetable*}